\input epsf

\documentclass[10pt]{amsart}

\usepackage{amscd, amsmath, amsthm}

\def\mylabel#1{\label{#1}   \proplabeL{#1} \hskip-3pt  }

\def\mylabel#1{\label{#1}}

\def\pd{\partial}

\newtheorem{theorem}{Theorem}[section]

\newtheorem{proposition}[theorem]{Proposition}
\newtheorem{conjecture}[theorem]{Conjecture}

\theoremstyle{definition}     
\newtheorem{definition}[theorem]{Definition}

\theoremstyle{remark}

\numberwithin{equation}{section}

\newcommand{\simrightarrow}{\smash{\mathop{\rightarrow}\limits^{\sim}}} 

\begin{document}

\title[ Central charges, Symplectic forms, and hypergeometric series ]{ 
{Central charges, symplectic forms, and hypergeometric series 
in local mirror symmetry}
}

\author[S. Hosono]{Shinobu Hosono}

\address{
Graduate School of Mathematical Sciences, 
University of Tokyo, Komaba Meguro-ku, 
Tokyo 153-8914, Japan
}
\email{hosono@ms.u-tokyo.ac.jp}


\begin{abstract}
We study a cohomology-valued hypergeometric series which naturally 
arises in the description of (local) mirror symmetry. 
We identify it as a central charge formula for BPS states and study its 
monodromy property from the viewpoint of Kontsevich's homological 
mirror symmetry. In the case of local mirror symmetry, we will identify 
a symplectic form, and will conjecture an integral and symplectic 
monodromy property of a relevant hypergeometric series of 
Gel'fand-Kapranov-Zelevinski type.
\end{abstract}

\maketitle


\centerline{\bf Table of Contents }

{\S 1.} Introduction  

\vskip0.2cm
{\S 2.} Central charge formula in terms of $w(x;\frac{J}{2\pi i})$

\vskip0.1cm
{\S 3.} Local mirror symmetry I -- $X=\widehat{\mathbf C^2/\mathbf Z_{\mu+1}}$

\hskip0.4cm
(3-1) Mirror symmetry and hyperk\"ahler rotation

\hskip0.4cm
(3-2) GKZ hypergeometric series

\hskip0.4cm
(3-3) Example ($\mu=2$)

\vskip0.2cm
{\S 4.} K. Saito's differential equations 

\hskip0.4cm
(4-1) K. Saito's system for a primitive form $\mathcal U_0(a)$

\hskip0.4cm
(4-2) GKZ system for a primitive form $\mathcal U(a)$

\vskip0.2cm
{\S 5.} Central charge formula and $G$-Hilb

\vskip0.1cm
{\S 6.} Local mirror symmetry II -- $X=\widehat{\mathbf C^3/G}$

\hskip0.4cm
(6-1) Period integrals and GKZ systems

\hskip0.4cm
(6-2) Central charge formula and symplectic forms

\hskip0.4cm
(6-3) Prepotential and a symplectic D-brane basis 

\hskip0.4cm
(6-4) Examples

\vskip0.2cm
{\S 7.} Conclusion and discussions

\vskip0.3cm

{Appendix A.} Integrating over vanishing cycles

{Appendix B.} Differential equation (A.9)

\vskip0.5cm

\noindent
\section{ {\bf Introduction } }

Let us consider a (famous) hypergeometric series of one variable 
$[1,x] \in \mathbf P^1$;
\begin{equation}
w(x)=\sum_{n \geq 0} \frac{(5n)!}{(n!)^5} x^n 
\mylabel{eqn:w0}
\end{equation}
This is a hypergeometric series of type $\,_4F_3(
\frac{1}{5},\frac{2}{5},\frac{3}{5},\frac{4}{5};1,1,1;x)$ which arises 
in the mirror symmetry of quintic hypersurface $X_5 \subset \mathbf P^4$. 
This hypergeometric series 
(\ref{eqn:w0}) represents one of the period integrals of the mirror 
quintic $X_5^\vee$ and satisfies the following differential equation 
(Picard-Fuchs) equation:
\begin{equation}
\{ \theta_x^4 -5^5 x 
(\theta_x+\frac{4}{5})
(\theta_x+\frac{3}{5})
(\theta_x+\frac{2}{5})
(\theta_x+\frac{1}{5}) \} w(x) =0\;,
\end{equation}
where $\theta_x:=x \frac{d \; }{dx}$. 
See the original work by Candelas et al \cite{CdOGP} for the description of 
the mirror family and its period integrals. 

As it is clear in the form of differential equation, the 
regular singularity at $x=0$ has a distinguished property, i.e., the monodromy 
around this point is {\it maximally unipotent} \cite{Mor}. 
In physics, the point $x=0$ is called a {\it large complex structure 
limit} and plays important roles, e.g., near this point, we 
evaluate the quantum corrections to the classical geometry of 
the $\sigma$-model on the quintic $X_5$. 
Here we focus on the construction of local solutions 
about $x=0$ by the classical Frobenius method; 
\begin{eqnarray*}
&w_0(x):=w(x) \;,\; 
w_1(x):=
\frac{\partial \;}{\partial \rho} w(x;\rho)\vert_{\rho=0} \;,\; \\
&w_2(x):=
\frac{\partial^2\;}{\partial \rho^2} w(x;\rho)\vert_{\rho=0} \;,\;
w_3(x):=
\frac{\partial^3 \;}{\partial \rho^3} w(x;\rho)\vert_{\rho=0} \;,\;
\mylabel{eqn:Frob}
\end{eqnarray*}
where $w(x;\rho):=\sum_{n \geq0} 
\frac{\Gamma(1+5(n+\rho))}{\Gamma(1+(n+\rho))^5} x^{n+\rho}$. 
With the mirror symmetry of $X_5$ in $\mathbf P^4$ and $X_5^\vee$ 
in mind, we introduce the following cohomology-valued 
hypergeometric series;
\begin{equation}
w(x;\frac{J}{2\pi i}):=
w(x)+
w_1(x)(\frac{J}{2\pi i}) + 
w_2(x)(\frac{J}{2\pi i})^2 + 
w_3(x)(\frac{J}{2\pi i})^3 \;, 
\mylabel{eqn:w(x;J)}
\end{equation}
where $J$ is the ample, integral generator of Pic$(X_5)=H^{1,1}(X_5) \cap 
H^2(X_5,\mathbf Z)$. In this form, we note that the classical Frobenius 
method is concisely summarized as the Taylor expansion 
$w(x;\rho)|_{\frac{J}{2\pi i}}$ with respect to the nilpotent element 
$J$. Although this seems just an advantage in  bookkeeping, the following 
observation reported in \cite{Hos} indicates that we have 
{\it more than} that in (\ref{eqn:w(x;J)}): 
 
\vskip0.3cm
\noindent
{\bf Observation:} {\it 
Arrange the Taylor series expansion of the cohomology-valued 
hypergeometric series $w(x;\frac{J}{2\pi i})$ as 
\begin{equation}
w(x;\frac{J}{2\pi i})=w^{(0)}(x) + 
w^{(1)}(x)(J - \frac{c_2(X_5) J}{12} - \frac{11}{2}\frac{J^2}{5})
+ w^{(2)}(x)\frac{J^2}{5} + 
w^{(3)}(x)(-\frac{J^3}{5}). 
\mylabel{eqn:quintic}
\end{equation}
Then the monodromy matrices of the coefficient hypergeometric series 
$w^{(0)}(x)$, $w^{(1)}(x)$, $w^{(2)}(x)$, $w^{(3)}(x)$ are integral and 
symplectic (with respect to the symplectic form of the mirror, see below). }
\vskip0.3cm

The integral and symplectic properties of the solutions 
$w^{(k)}(x)\; (k=0,1,2,3)$ stem from those of the middle homology 
group of the mirror $H_3(X_5^\vee,\mathbf Z)$. The point here is 
that we can recover these integral, symplectic properties from the 
series $w(x;\frac{J}{2\pi i})$ through suitably  
arranging a basis of $H^{even}(X,\mathbf Q)$ near the large 
complex structure limit. Since there is a symplectic, 
integral structure on $H^{even}(X,\mathbf Q)$ which comes from 
$K(X)$, the Grothendieck group of algebraic vector bundles on $X$, 
it is natural to conjecture (Conjecture 2.2) that the cohomology-valued 
hypergeometric series describes a ``pairing'' between $K(X)$ 
and $H_3(X_5^\vee,\mathbf Z)$, as proposed in \cite{Hos}.

\vskip0.4cm

The aims (and main results) of this paper are, 1) to make an interpretation 
of the cohomology-valued hypergeometric series from the viewpoint of 
homological mirror symmetry, and to give a general definition of 
the so-called central charge of BPS states 
({\bf Conjecture 2.2}, {\bf Conjecture 6.3}), 
2) to present supporting evidences for the interpretation in the cases 
from local mirror symmetry in dimensions two and three 
(Section 5, Section 6). 
\vskip0.4cm

The idea of the local mirror symmetry is simply to focus on the 
Calabi-Yau geometry near rigid curves or (Fano) surfaces in 
a Calabi-Yau manifold \cite{CKYZ}. 
Recently this idea has been providing us useful 
testing grounds for several duality symmetries proposed in string theory. 
In particular, when the relevant Calabi-Yau geometry is toric, remarkable 
progresses have been made (, see \cite{AKMV} and references therein). 
In this paper, for the local Calabi-Yau geometries, we will consider the 
crepant resolutions 
$\widehat{\mathbf C^2/\mathbf Z_{\mu+1}}$ of the two dimensional 
canonical singularity and also $\widehat{\mathbf C^3/G}$ with finite abelian 
group $G \subset SL(3,\mathbf C)$. Although we restrict our attentions to 
these cases, our arguments should extend to more general 
local Calabi-Yau geometries.

\vskip0.4cm

When studying relevant Gel'fand-Kapranov-Zelevinski(GKZ) 
hypergeometric series, we will find its connection to the primitive forms 
introduced by K.Saito in the deformation theory of singularities 
({\bf Proposition 4.2}, {\bf Proposition 4.4}). 
This seems to be interesting in its own right, since GKZ hypergeometric 
series may provide a simple way to express the `period integrals' of 
the primitive forms (or oscillating integrals) in the theory of singularities.

\vskip0.5cm

The construction of this paper is as follows: In section 2, we give 
a definition of the central charge assuming Kontsevich's homological 
mirror symmetry. Following the observations made in \cite{Hos}, we interpret 
the cohomology-valued hypergeometric series as the central charge for 
the cases of $X$ compact. In section 3, we consider the local mirror 
symmetry of $X=\widehat{\mathbf C^2/\mathbf Z_{\mu+1}}$ and study 
in detail the GKZ hypergeometric series in this case. In section 4, 
we connect the GKZ system to K.Saito's system of differential 
equations which describes the deformation  of singularities.  
In section 5, we extract a (homological) mirror map 
$mir: K^c(X) \simrightarrow H_2(Y,\mathbf Z)$ 
from our interpretation of the cohomology-valued 
hypergeometric series, and find a consistency with the mirror symmetry 
in terms of the hyperk\"ahler rotation. In section 6, we formulate 
our conjecture on cohomology-valued hypergeometric series 
for three dimensional cases $X=\widehat{\mathbf C^3/G}$. 
We also find a closed formula 
for the prepotential. Two explicit examples are presented to 
support our conjecture. In Appendix A and B, we evaluate period 
integrals over vanishing cycles for the mirror geometry of 
$X=\widehat{\mathbf C^3/\mathbf Z_3}$. There, we see that 
a certain noncompact vanishing cycle arises in the description of 
the isomorphism $mir: K^c(X) \simrightarrow H_3(Y,\mathbf Z)$. 
The relation to the period integral of a primitive form is 
also elucidated.

\vskip0.3cm
\noindent
{\bf Acknowledgments:} The author would like to thank T. Terasoma for 
discussions on the GKZ hypergeometric system. He also would like to thank 
K. Oguiso and S.-T.Yau for valuable comments on this work. 

\vfill\eject
\vskip0.7cm

\section{ {\bf Central charge formula in terms of $w(x;\frac{J}{2\pi i})$ }}

Here, we will interpret the cohomology-valued hypergeometric series 
from the viewpoint of Kontsevich's homological 
mirror symmetry \cite{Ko}. 

Let $X$ be a Calabi--Yau 3 fold and $Y$ be a mirror of $X$. Following 
Kontsevich, we consider the bounded derived category 
$D^b(Coh(X))$ of coherent sheaves (D-branes of B type) on $X$. 
On the other hand, for the mirror side, we 
consider the derived Fukaya category $D Fuk(Y,\beta)$ with the K\"ahler 
form viewed as a symplectic form $\beta$. The objects of the latter 
category consist of (graded) Lagrangian submanifolds with a flat $U(1)$ 
bundle on each of them (D-branes of type A) 
and morphisms are given by the Floer 
homology for Lagrangian submanifolds, and this category forms 
a triangulated category (see \cite{FO3} for a more precise definition). 
Kontsevich proposed that these two different categories are equivalent (as 
triangulated categories) when $X$ and $Y$ are mirror symmetric, and 
also that this should be a mathematical definition of mirror symmetry. This 
conjecture itself is of great interest; however let us consider 
this conjecture at a more tractable level, i.e. at the level of cohomology 
or K-groups as shown in the second line below;
\begin{equation}
\begin{CD}
D^b(Coh(X)) @>{Mir}>> D Fuk(Y,\beta) \\
@VVV   @VVV \\
K(X) \text{ or } H^{even}(X,\mathbf Q) @>{mir}>> H_3(Y,\mathbf Z) \\
\end{CD}
\mylabel{eqn:Cdiag}
\end{equation}
where the left vertical arrow represents the natural map 
from $D^b(Coh(X))$ to the 
K-group of algebraic vector bundles $K(X)$, and its composition with the 
Chern character homomorphism $ch(\cdot)$ if we further map to $H^{even}(X, 
\mathbf Q)=\oplus_{p=0}^3 H^{2p}(X,\mathbf Q)$. The right vertical arrow is 
given simply by taking the homology classes of the graded Lagrangian cycles. 
In the second line, the equivalence, $Mir$, of the two categories 
becomes simply an isomorphism, 
$mir$, between the K-group and $H_3(Y,\mathbf Z)$. We should note that 
this is not simply an isomorphism but an isomorphism with the symplectic 
structures, i.e.
$$
mir: \;(K(X), \chi(E,F)) \;\; \simrightarrow \;\; 
(H_3(Y,\mathbf Z),\#(L_E\cap L_F) ), 
$$
where $\chi(E,F)=\int_X ch(E^\vee)ch(F)Todd_X$ and $\#(L_E\cap L_F):=
\int_Y \mu_{L_E} \cup \mu_{L_F}$ with the Poincar\'e duals $\mu_{L_E}, 
\mu_{L_F} \in H^3(Y,\mathbf Z)$ of the mirror homology cycles 
$L_E:=mir(E), L_F:=mir(F)$. Here we remark that the Euler number $\chi(E,F)$ 
is anti-symmetric due to Serre duality and $K_X=0$, and also non-degenerate. 
Thus $\chi(E,F)$ introduces a symplectic structure on $K(X)$, which is the 
mirror of the symplectic structure on $H_3(Y,\mathbf Z)$.  

In the diagram (\ref{eqn:Cdiag}), we assume that a complex structure is fixed 
to define $D^b(Coh(X))$.  Correspondingly the symplectic form 
(K\"ahler form) $\beta$, determined by the mirror map, is fixed in the right 
hand side. On the other hand, we may change the 
(complexified) K\"ahler class of $X$ which corresponds to the complex 
structure moduli of $Y$ under the mirror map. Changing the 
(complexified) K\"ahler structure amounts to changing the polarization 
and thus results in varying the stability condition on the sheaves on $X$. 
This change of the stability ($\Pi$-stability) condition has been studied 
in \cite{Do} as a stability of BPS D-branes 
and its mathematical aspects are elaborated in \cite{Br}. Here, we will 
not go into the detailed definition of $\Pi$-stability, but we will 
propose a closed formula for the {\it central charge} which is 
indispensable for an explicit description of $\Pi$-stability. 
 
\begin{definition} {\bf (Central charge formula.)} 
Assume $K(X)$ is torsion-free, and 
let $E_1,\cdots,E_r$ be a $\mathbf Z$-basis of $K(X)$. Let $\Omega(Y_x)$ 
be a holomorphic 3-form of the mirror family $\{Y_x\}_{x \in \mathcal B}$ of 
$X$ with $x=(x_1,\cdots,x_r)$ being the local parameter near a large 
complex structure limit. Denote by ${\mathbf C}\{x\}[\log x]$ the 
polynomial ring of $\log x_1,\cdots,\log x_r$ over the ring ${\mathbf C}\{x\}$ 
of convergent power series. 
Under the mirror symmetry (\ref{eqn:Cdiag}), we define the following 
$\mathcal Z_x$ as an element in $K(X)\otimes \mathbf C\{x\}[\log x]$: 
\begin{equation}
{\mathcal Z}_x := \sum_{i,j} \int_{mir(E_i)} \Omega(Y_x) \; 
\chi^{ij} \; E_j^\vee  \; 
\mylabel{eqn:calZ}
\end{equation} 
with $(\chi^{ij}):=(\chi(E_i,E_j))^{-1}$. Then the central charge of 
$F \in K(X)$ is defined by  
\begin{equation}
Z_t(F)=\int_X ch(F) \; ch({\mathcal Z}_x) \; Todd_X\;\;,
\end{equation}
where $t=t(x)$ (or $x=x(t)$) is the mirror map connecting the local 
parameters $x_1,\cdots,x_r$ to those $t_1,\cdots,t_r$ of 
the (complexified) K\"ahler moduli space of $X$. 
\end{definition}

In the above definition, it should be noted that 
$\mathcal Z_x$ does not depend on the choice of a basis $E_1,\cdots,E_r$. 
Also the central charge $Z_t(F)$ contains full 'quantum corrections' 
as a function of $t_1,\cdots,t_r$ ( cf. the asymptotic formula given 
in \cite{Do}). 

Let us connect our hypergeometric series $w(x;\frac{J}{2\pi i})$ to 
the central charge above. Before doing this, we remark that, in the mirror 
symmetry of Calabi--Yau hypersurfaces due to Batyrev\cite{Ba1}, 
the hypergeometric series (\ref{eqn:w0}) is naturally generalized 
to the Gel'fand--Kapranov--Zelevinski (GKZ) 
hypergeometric series of multi-variables 
$x_1,\cdots,x_r$ \,\cite{GKZ1},\cite{Ba2}.
Using the GKZ hypergeometric series, and also suitable integral, 
(semi-)ample generators $J_1,\cdots,J_r$ of $H^{1,1}(X)\cap H^2(X,\mathbf Z)$, 
we have a cohomology-valued hypergeometric series $w(x;\frac{J}{2\pi i})$ as 
a generalization of (\ref{eqn:w(x;J)}) (see Sect.2 of \cite{Hos} for 
more details).

\begin{conjecture} \mylabel{thm:conj}
The cohomology-valued hypergeometric series 
(\ref{eqn:w(x;J)}) gives the expression 
$ch(\mathcal Z_x) \in H^{even}(X)\otimes{\mathbf C}\{x\}[\log x]$;
\begin{equation}
w(x_1,\cdots,x_r;\frac{J_1}{2\pi i},\cdots,\frac{J_r}{2\pi i})=
\sum_{i,j} \int_{mir(E_i)} \Omega(Y_x) \; \chi^{ij} \; ch(E_j^\vee) \;.
\mylabel{eqn:chargeWxJ}
\end{equation}
Using this, and also the mirror map $t=t(x)$, we can write the central 
charge $Z_t(F)$ of $F \in K(X)$ as
\begin{equation}
Z_t(F)=\int_X ch(F)\; w(x;\frac{J}{2\pi i}) \; Todd_X \;\;.
\end{equation}
\end{conjecture}

Here we note that the hypergeometric series has a finite radius of 
convergence and shows a monodromy property when it is analytically 
continued around its (regular) singularities. As noticed by Kontsevich, 
this monodromy property should be mirrored to some linear (symplectic) 
transformations on $ch(E_i)$ which come from Fourier--Mukai transforms 
on $D^b(Coh(X))$. 
If we postulate that the cohomology-valued hypergeometric series has an 
invariant meaning under these monodromy actions, our cohomology-valued 
hypergeometric series $w(x;\frac{J}{2\pi i})$ provides a connection 
between these two different `monodromy' transforms on the two sides. 
The conjectural formula (\ref{eqn:chargeWxJ}) has been tested
in the case $X$ is an elliptic curve, (lattice polarized) K3 surfaces, and 
several Calabi--Yau hypersurfaces in \cite{Hos}. Cohomology-valued 
hypergeometric series are utilized also in \cite{Gi},\cite{LLY},\cite{Sti} 
in a slightly different form. We remark that, for our Conjecture 
\ref{thm:conj} to work, the definition given in \cite{Hos} is crucial.

As studied in \cite{Mu} for the cases of K3 surfaces and abelian varieties, 
and in \cite{Or} for general, 
the Fourier--Mukai transform is a self-equivalence of 
the category $D^b(Coh(X))$ which takes the form
$$
\Phi^{\mathcal P}(\cdot)=\mathbf R{p_2}_* ( p_1^*(\cdot) 
 \otimes^{\hskip-7pt \,^\mathbf L} \mathcal P)
$$
where $\mathcal P$ is an object in $D^b(Coh(X\times X))$, 
called the {\it kernel}, 
and $p_1$ and $p_2$ are, respectively, the natural projections to the 
first and the second factor from $X \times X$. Due to a result in \cite{Or}, 
we may always assume the above form, i.e., there exists a suitable kernel 
$\mathcal P$, for 
any equivalence $\Phi: D^b(Coh(X)) \simeq D^b(Coh(X))$ as triangulated 
category. It is rather easy to see that the monodromy transforms around 
the large complex structure limit are given by tensoring invertible sheaves, 
which may be expressed by the kernels;
$$
\mathcal P : \; \cdots \rightarrow 0 \rightarrow \mathcal 
O_\Delta \times p_2^*(\mathcal O_X(D)) \rightarrow 0 \cdots \;\;,
$$
with $D \in Pic(X)$ and $\Delta$ representing the diagonal in 
$X\times X$. Kontsevich predicted that a monodromy transform 
associated to a vanishing cycle, the Picard--Lefschetz transform, 
has its mirror FM transform with its kernel,
$$
\mathcal P: 
\; \cdots \rightarrow 0 \rightarrow \mathcal O_{X\times X} 
\rightarrow \mathcal O_X \rightarrow 0 \cdots \;\;.
$$
Seidel and Thomas \cite{ST} (and Horja \cite{Hor}) 
generalized the above kernel, 
associating it to so-called spherical objects 
$\mathcal E \in D^b(Coh(X))$ with the defining property: 
$Ext^i(\mathcal E,\mathcal E)=0 \;(i\not= 0,n), 
\,= \mathbf C \;(i=0,n)$ where $n=\text{dim}\, X$.  
For each spherical object, we have a kernel given by the mapping cone;
$$
\mathcal P= Cone( \mathcal E^\vee \otimes^{\mathbf L} \mathcal E 
\rightarrow \mathcal O_\Delta) \;\;.
$$ 
The equivalence $\Phi^\mathcal P$ is called a Seidel-Thomas twist. 
We will see these equivalences in the corresponding monodromy property 
of our hypergeometric series.

\vskip1cm
\section{ {\bf Local mirror symmetry I --- 
$X=\widehat{\mathbf C^2/\mathbf Z_{\mu+1}}$ }}

In this section and the subsequent sections, we will test our Conjecture 
\ref{thm:conj} for the case of mirror symmetry of non-compact toric 
Calabi-Yau manifolds (local mirror symmetry). Batyrev's idea of mirror 
symmetry \cite{Ba1} still makes sense for such non-compact toric 
Calabi-Yau manifolds as well as the attractive proposal by 
Strominger-Yau-Zaslow(SYZ)\cite{SYZ}, which is closely 
related to the homological mirror symmetry (\ref{eqn:Cdiag}). 
Mirror symmetry of non-compact toric Calabi-Yau manifolds 
and also Fano varieties are formulated in a language of  
Landau-Ginzburg theory in \cite{HIV}.

\vskip0.5cm
{\bf (3-1) Mirror symmetry and hyperk\"ahler rotation.} 
Let us consider the minimal resolution of a two dimensional simple  
singularity; $X=\widehat{\mathbf C^2/\mathbf Z_{\mu+1}}$. This is an example 
of two dimensional, non-compact, toric Calabi-Yau manifold. Two dimensional 
Calabi-Yau manifolds are hyperk\"ahler, and it is known that 
the mirror symmetry of them is well-understood by the hyperk\"ahler 
rotation, see e.g. \cite{GW}\cite{Huy}. 
Our minimal resolution $X$ has a natural 
hyperk\"ahler structure, and therefore its mirror is $X$ itself with a 
different complex structure after a suitable rotation. To describe the 
mirror symmetry, let us first write the quotient 
$\mathbf C/\mathbf Z_{\mu+1}$ by a hypersurface 
$UV=W^{\mu+1}$ in $\mathbf C^3$. Bowing up the singularities at the 
origin $\mu$ times results in the minimal resolution $X$, and thereby 
we introduce exceptional curves $C_i\cong \mathbf P^1$ $(i=1,\cdots,\mu)$. 
On the other hand, we may deform the defining equation $UV=W^{\mu+1}$ 
to $UV=a_1+a_2W+\cdots+a_{\mu+2}W^{\mu+1}$ with introducing finite sizes 
to vanishing cycles $L_i\approx S^2$ $(\mu=1,\cdots,\mu)$. Note that 
the number of vanishing cycles are given by the Milnor number 
$\mu=\text{dim} R_J$, where $R_J$ is the Jacobian ring of the singularity 
$UV=W^{\mu+1}$. The vanishing cycles are Lagrangians, 
and become holomorphic cycles under a suitable 
hyperk\"ahler rotation. The holomorphic geometry after the rotation 
is bi-holomorphic to the blown-up geometry of $X$. If we forget about the 
role of the $B$-fields, this describes the mirror symmetry of $X$. (See e.g. 
\cite{Huy} and references therein   
for full details of the mirror symmetry via the hyperk\"ahler rotation.) 
Here we note that the intersection form of these vanishing 
cycles are given in both holomorphic and Lagrangian geometry by 
$$
(C_i\cdot C_j)=(\# L_i\cap L_j)=-\mathcal C_{ij}\;\;,
$$
where $\mathcal C_{ij} \,(1\leq i,j \leq \mu)$ is the Cartan matrix for 
the root system of $A_{\mu}$.

\vskip0.5cm
{\bf (3-2) GKZ hypergeometric series}. The minimal resolution $X=
\widehat{\mathbf C^2/\mathbf Z_{\mu+1}}$ is a (non-compact) toric 
variety whose resolution is described by a two dimensional fan 
$\Sigma$ with its integral generators $\nu_i$ for one dimensional cones 
(see Fig.1). We set 
$$
A=\{ \nu_1,\nu_2,\cdots,\nu_{\mu+2}\}=
\{\;(1,0),(1,1),\cdots,(1,\mu+1) \;\} \;\;.
\mylabel{eqn:A}
$$
The half-lines $\overline{o\nu}_i$ $(i=1,\cdots,\mu+2)$ 
from the origin $o=(0,0)$ 
constitute the one dimensional cones of the resolution diagram $\Sigma$. 
In Batyrev's mirror symmetry, the resolution diagram of $X$, up to flop 
operations, is identified with the Newton polytope of the defining equation of 
its mirror $Y$. This construction has been extended to the cases of 
non-compact toric Calabi-Yau maniflods in \cite{CKYZ}\cite{HIV}, and for 
$X=\widehat{\mathbf C^2/\mathbf Z_{\mu+1}}$, the mirror $Y$ 
is given by $U^2+V^2+f_\Sigma(W)=0 \subset \mathbf C^2 \times \mathbf C^*$ 
with
$$
f_\Sigma(W)= a_1 + a_2 W^1 + a_3 W^2 + \cdots + a_{\mu+2} W^{\mu+1} \;,
$$
where $(U,V,W) \in \mathbf C^2 \times \mathbf C^*$. 
In the case of local mirror symmetry, the meaning of the period 
integrals of holomorphic two form becomes less clear than the 
compact cases. Here, motivated by \cite{CKYZ}, 
we consider the following integral 
\begin{equation}
\Pi_\gamma(a):=\frac{i}{2\pi^2} \int_\gamma 
\frac{1}{U^2+V^2+f_\Sigma(a;W)} dU dV \frac{dW}{W} \;\;,
\mylabel{eqn:Pi}
\end{equation}
for each cycle in the complement of the hypersurface  
$$
\gamma \in H_3(\mathbf C^2\times\mathbf C^*
\setminus Y,\mathbf Z). 
$$
Note that we have an isomorphism $H_3(\mathbf C^2\times\mathbf C^* 
\setminus Y, \mathbf Z) \simeq H_2(Y,\mathbf Z)$ for the hypersurface 
$Y \subset \mathbf C^2 \times \mathbf C^*$ 
(see P.46 of \cite{Di} for example). 

In the next section, we will relate this `period integral' to  
K. Saito's primitive form (or Gel'fand-Leray form) in the deformation 
theory of singularities \cite{Sa}\cite{AGV}.  
Here we set up a hypergeometric differential 
equations (GKZ system) satisfied by $\Pi_\gamma(a)$. This 
GKZ hypergeometric system 
is also referred to as $A$-hypergeometric system since it is 
described by the set 
$A$ through the {\it lattice of relations},
\begin{equation}
\mathcal L=\{\;(l_1,l_2,\cdots,l_{\mu+2}) \,|\, l_1\nu_1+l_2\nu_2+\cdots
+l_{\mu+2}\nu_{\mu+2}=(0,0) \;\}.
\mylabel{eqn:latticeL}
\end{equation}
Using this lattice of relations, the system is written as 
\begin{equation}
\Box_l \Pi_\gamma(a)=0\;(l\in \mathcal L)\;\;,\;\;
\mathcal Z_i \Pi_\gamma(a) =0 \;(i=1,2) \;,
\mylabel{eqn:GKZdiff}
\end{equation}
where 
$$
\Box_l=\left(\frac{\pd\;}{\pd a}\right)^{l^+}-
\left(\frac{\pd\;}{\pd a}\right)^{l^-}\;\;,\;\;
\left(
\begin{matrix}
\mathcal Z_1 \\
\mathcal Z_2 \\  \end{matrix} \right)
=
\left(
\begin{matrix}
\theta_1+\theta_2+\cdots 
+\theta_{\mu+2} \\
\theta_2+2 \theta_3\cdots+(\mu+1)\theta_{\mu+2} \\
\end{matrix} \right)
$$
with $l=l^+-l^-$ ($l_i^\pm:=\frac{1}{2}(l_i\pm |l_i|)$ 
and $\theta_k=a_k\frac{\pd\;}{\pd a_k}$. From the formal 
solutions of this system in \cite{GKZ1}, it is easy to write down our  
$w(x;\frac{J}{2\pi i})$ (, see  (3-3) below). 
An important aspect of this system 
is that there is a natural toric compactification $\mathcal M_{Sec(\Sigma)}$ 
of the parameter space 
$\{ (a_1,\cdots,a_{\mu+2}) \in (\mathbf C^*)^{\mu+2}/(\mathbf C^*)^2 \}$ 
in terms of the {\it secondary fan} $Sec(\Sigma)$ (, see \cite{GKZ2} and 
references therein), where the quotient by 
$(\mathbf C^*)^2$ is represented by the (scaling) 
operators  $\mathcal Z_1$ and $\mathcal Z_2$.  
This compactification plays an important role in the applications of mirror 
symmetry to Gromov-Witten invariants, since the large radius limit appears 
as an intersection point of the boundary toric divisors. Connecting the GKZ 
system to K. Saito's differential equations in singularity theory, we 
will see in section 4 that 
this compactification will also provide a natural 
way to compactify the deformation 
space of singularities which is local in nature (, see (4-1) for 
a brief description of the deformation space).

\vskip0.5cm
\centerline{\epsfxsize 10.5truecm\epsfbox{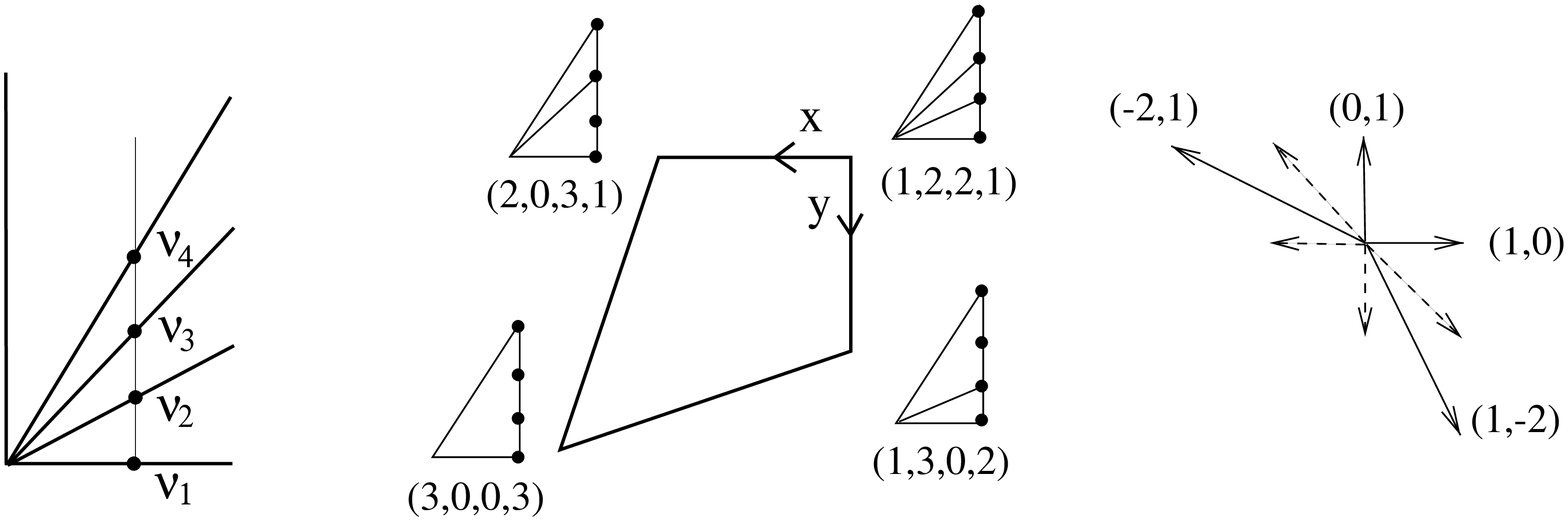}}

{\leftskip1cm \rightskip1cm 
\noindent 
{\sl {\bf Fig.1.} 
The resolution diagram (left), the secondary polytope (middle), 
and the secondary fan $Sec(\Sigma)$ (right) for $\mu=2$. The secondary 
polytope has its vertices parametrized by (regular) triangulations 
of the polytope as shown. For a triangulation $T$, the corresponding 
vertex is determined by a vector 
$v_T=(\varphi_T(\nu_1),\varphi_T(\nu_2), \varphi_T(\nu_3), \varphi_T(\nu_4))$ 
with $\varphi_T(\nu_i)=\sum_{\nu_i \prec \sigma }Vol(\sigma)$. As we 
see in this example ($\mu=2$), the convex hull of these vertices lies on 
$\mathcal L_{\mathbf R} + v_T$ with a vector $v_T$, where 
$\mathcal L_{\mathbf R}=\mathcal L\otimes\mathbf R$. 
Normal cones of the secondary polytope determines the secondary fan.
\par}}

\vskip0.5cm
{\bf (3-3) Example ($\mu=2$)}. Here we present some explicit calculations  
of cohomology-valued hypergeometric series for the case $\mu=2$, 
instead of giving general formulas valid for any $\mu$. In this case, 
the cohomology-valued hypergeometric series is simple and takes 
the following form;
$$
w(x;y;\frac{J_1}{2\pi i}, \frac{J_2}{2\pi i})=
w(x;y;\rho_1,\rho_2)|_{\rho_1=\frac{J_1}{2\pi i}, \rho_2=\frac{J_2}{2\pi i}}\;,
$$
where 
$w(x;y;\rho_1,\rho_2)=\sum_{n,m\geq 0} c(n+\rho_1,m+\rho_2)
x^{n+\rho_1}y^{m+\rho_2}$ with 
$$
c(n,m)=1/\big(\Gamma(1+n)\Gamma(1-2n+m)\Gamma(1+n-2m)\Gamma(1+m)\big) \;.
$$
$J_1$ and $J_2$ are semi-ample classes which are dual to the exceptional 
curves $C_1$ and $C_2$ (, i.e. the toric divisors 
$D_{\nu_2}$ and $D_{\nu_3}$), 
respectively. The local parameters $x:=\frac{a_1a_3}{a_2^2}, 
y:=\frac{a_2a_4}{a_3^2}$ are depicted in Fig.1. 
Here we remark that the secondary polytope in Fig.1 sits on 
a translation of $\mathcal L_{\mathbf R}$, i.e. 
$\mathcal L_{\mathbf R}+v_T$, and the 
summation in $\sum_{n,m\geq 0}c(m,m)x^m y^n$ is in fact that 
over the integral points inside the dual of the normal cone from the 
vertex $v_T=(1,2,2,1)$. We may recognize this fact in the relation 
$x^ny^m=a^{n l^{(1)} + ml^{(2)} }$, where by definition 
$l^{(1)}=(1,-2,1,0)$ and $l^{(2)}=(0,1,-2,1)$ generate the dual of 
the normal cone form the vertex $v_T$. We remark also that 
the coefficient $c(n,m)$ are determined from the entries of 
$n l^{(1)}+m l^{(2)}$;
$$
(n \;m)\left( 
\begin{matrix}
1 & -2 & 1  & 0  \\
0 &  1 & -2 & 1 \\ 
\end{matrix}\right)
= \left(
\begin{matrix}
n & -2n &  n  & 0  \\
0 &  m  & -2m & m \\ 
\end{matrix}\right) \;\;,
$$
where the row vectors determine the arguments 
of the gamma functions in the denominator of $c(n,m)$. In this form, 
it is rather clear how the formula generalizes to arbitrary $\mu$ (, see 
\cite{CKYZ},\cite{dOFS} for example). 

Since the 
ring $H^{even}(X,\mathbf Z)$ is generated by $1, J_1, J_2$, 
we have the expansion; 
$$
w(\vec x,\frac{\vec J}{2\pi i})=1+w_1(x,y) J_1 +w_2(x,y) J_2 \;\;,
$$
with 
$w_1(x,y)=\frac{1}{2\pi i} \log x + \cdots$, 
$w_2(x,y)=\frac{1}{2\pi i} \log y + \cdots$. 
The mirror map is defined from the relations,
\begin{equation}
q_1:=e^{2\pi i w_1(x,y)}=x(1+g_1(x,y)) \;\;,\;\;
q_2:=e^{2\pi i w_2(x,y)}=y(1+g_2(x,y)) \;\;,
\mylabel{eqn:qs}
\end{equation}
where $g_1(x,y),g_2(x,y)$ represent powerseries of $x$ and $y$.  Then 
$t_1:=\frac{1}{2\pi i} \log q_1(=w_1(x,y))$ and 
$t_2:=\frac{1}{2\pi i}\log q_2(=w_2(x,y))$ are the 
complexified K\"ahler moduli and measure the volumes of the exceptional 
curves $C_1$, and $C_2$, respectively. The inverse relation $x=x(q_1,q_2), 
y=y(q_1,q_2)$ of (\ref{eqn:qs}) is often referred to as the mirror map, and 
has the following properties (see Proposition 4.4 for a proof);

\vskip0.5cm
\begin{proposition} \mylabel{thm:Mmap}
\item[1)] The mirror map $x=x(q_1,q_2), y=y(q_1,q_2)$ is rational 
and is expressed by $x=\frac{a_1a_3}{a_2^2}, 
y=\frac{a_2a_4}{a_3^2}$ with $a_i$'s determined through 
\begin{equation}
a_1+a_2W+a_3W^2+a_4W^3=(1+W)(1+q_1W)(1+q_1q_2W) \;\;.
\mylabel{eqn:qWprod}
\end{equation}
Concretely, it has the form;
\begin{equation}
x=\frac{q_1(1+q_2+q_1q_2)}{(1+q_1+q_1q_2)^2}\;,\;
y=\frac{q_2(1+q_1+q_1q_2)}{(1+q_1+q_1q_2)^2}\;\;.
\mylabel{eqn:mirrormap}
\end{equation}
\item[2)] The discriminant of the GKZ system (\ref{eqn:GKZdiff})  
consists of three components; $x=0, y=0$ and $dis(x,y)=0$ with
$$
dis(x,y)=1-4x-4y+18xy-27x^2y^2=
\frac{(1-q_1)^2(1-q_2)^2(1-q_1q_2)^2}{(1+q_1+q_1q_2)^2(1+q_2+q_1q_2)^2} \;.
$$
\end{proposition}

\vskip0.5cm
\centerline{\epsfxsize 11truecm\epsfbox{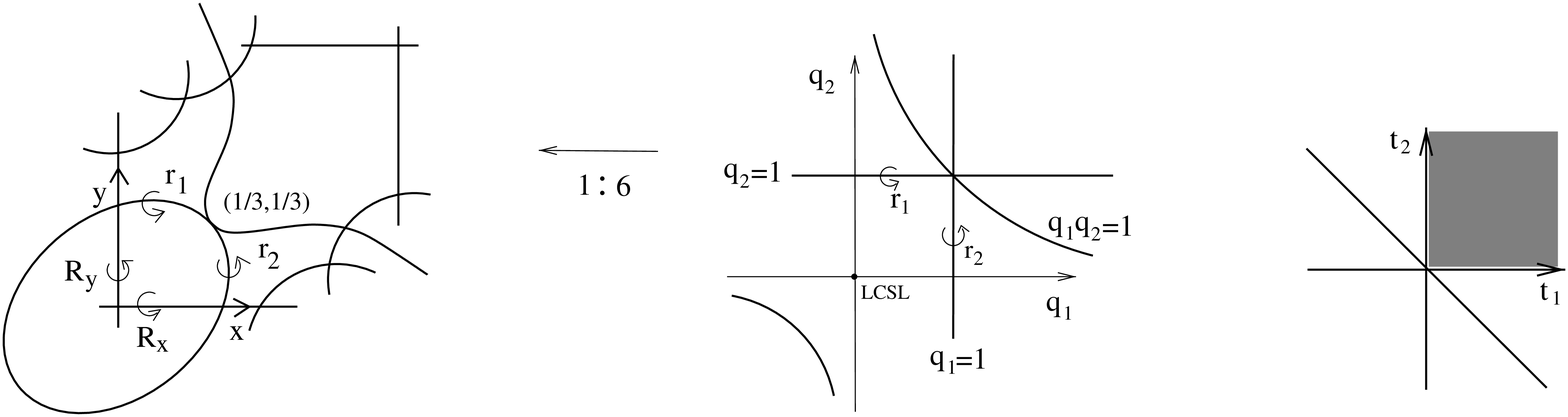}}

{\leftskip1cm \rightskip1cm 
\noindent 
{\sl {\bf Fig.2.} The discriminant $dis(x,y)=0$ in 
$\hat{\mathcal M}_{Sec(\Sigma)}$ (left), the mirror map (\ref{eqn:mirrormap}) 
from the $q_1q_2$-plane to the $xy$-plane (middle), 
and the complexified K\"ahler moduli $t_1,t_2$ 
with the complexified K\"ahler cone (right). The discriminant is an elliptic 
curve with a node at $(x,y)=(\frac{1}{3},\frac{1}{3})$, which corresponds to 
$(q_1,q_2)=(1,1)$. The mirror map is $6:1$ at generic $(q_1,q_2)$. Over 
the discriminant, it is $3:1$ and the inverse image of $dis(x,y)=0$ 
consists of the three lines $q_1=1, q_2=1,q_1q_2=1$ in the $q$-plane. 
\par}}

\vskip0.5cm
It is easy to see that $x=0, y=0$ are the toric boundary divisors whose 
intersection point define the large complex structure. Over the zeros of the 
discriminant $dis(x,y)$, we see vanishing cycles in $f_\Sigma(a,W)+U^2+V^2=0 
\subset \mathbf C^2\times \mathbf C^*$. In fact, in the holomorphic picture, 
$q_1=1$($q_2=1$) represents a vanishing volume limit of the exceptional 
curve $C_1$($C_2$). After a hyperk\"ahler rotation, these vanishing 
volumes are identified as the corresponding vanishing 
of the Lagrangian cycles $L_1, L_2$. 

We remark that the above 
Proposition \ref{thm:Mmap}, and also the interpretation made above 
generalize to arbitrary $\mu$ in a straightforward way. 
For example, the relation (\ref{eqn:qWprod}) should be read,  
\begin{eqnarray*}
&\sum_{k=0}^{k=\mu+1} 
a_{k+1} W^k &= (1+W)(1+q_1 W)(1+q_1q_2W)\cdots 
(1+q_1\cdots q_\mu W) \;\;, 
\end{eqnarray*}
and the discriminant is essentially given by the difference products the 
roots $1,-1/q_1$, $\cdots$, $-1/(q_1\cdots q_{\mu})$ (, see Proposition 4.1 
and 4.4).

\vskip0.5cm

In the rest of this subsection, we look at the mirror map 
(\ref{eqn:mirrormap}), and summarize its monodromy property. Let us first 
note that the moduli space $\mathcal M_{Sec(\Sigma)}$ is a two dimensional 
singular toric variety, and may be desingularized to 
$\hat{\mathcal M}_{Sec(\Sigma)}=Bl_4(\mathbf P^1 \times \mathbf P^1)$ 
after blowing up four points (, see the dashed lines in Fig.1). Then the 
discriminant $dis(x,y)=0$ describes a nodal elliptic curve written in Fig.2. 
Note that the mirror map form $(q_1,q_2)$ to $(x,y)$ is 
six to one at generic points 
because (\ref{eqn:mirrormap}) is invariant under the reflections;
\begin{equation}
r_1: q_1 \rightarrow {1}/{q_1}\;\;,\;\;
     q_2 \rightarrow q_1q_2 
\qquad,\qquad
r_2: q_2 \rightarrow q_1q_2 \;\;,\;\;
     q_1 \rightarrow {1}/{q_2} \;. 
\mylabel{eqn:reflections}
\end{equation}
which satisfy $r_1^2=r_2^2=(r_1r_2)^3=1$ and thus generate the symmetric 
group of order 3. By making an analytic continuation of the hypergeometric 
series, we can derive the above invariance group 
actions (\ref{eqn:reflections}) as the monodromy actions for the 
loops $r_1$ and $r_2$ depicted in Fig.2. Together with the monodromy 
matrices $R_x,R_y$ about 
the large complex structure limit, we summarize the monodromy generators 
with respect to a basis $\,^t(1,w_1(x,y),w_2(x,y))=
\,^t(1,\frac{1}{2\pi i}\log q_1,\frac{1}{2\pi i}\log q_2)$;
$$
r_1=
\left(\begin{matrix}
1 & 0 & 0 \cr
0 & -1 & 0 \cr
0 & 1 & 1 \cr \end{matrix} \right)\;,\;
r_2=
\left(\begin{matrix}
1 & 0 & 0 \cr
0 & 1 & 1 \cr
0 & 0 & -1 \cr \end{matrix} \right)\;;\;
R_x=
\left(\begin{matrix}
1 & 0 & 0 \cr
1 & 1 & 0 \cr
0 & 0 & 1 \cr \end{matrix} \right)\;,\;
R_y=
\left(\begin{matrix}
1 & 0 & 0 \cr
0 & 1 & 0 \cr
1 & 0 & 1 \cr \end{matrix} \right)\,.\;
$$
In general, we may state the monodromy property as follows;

\vskip0.3cm

\begin{proposition} \mylabel{thm:Uniform}
The mirror map (\ref{eqn:mirrormap}) uniformizes  
the solutions of the GKZ system (\ref{eqn:GKZdiff}) with the symmetric group 
$\mathcal S_\mu$ of order $\mu$, up to the shifts 
$w_i(x_1,\cdots,x_\mu)\mapsto w_i(x_1,\cdots,x_\mu)+1$ 
$(i=1,\cdots,\mu)$. In other words, the mirror coordinate $q_1,\cdots,q_\mu$ 
uniformizes the hypergeometric series with the symmetric group 
$\mathcal S_\mu$ of order $\mu$
\footnote{The author was pointed out by Prof. K.Saito that the monodromy group 
of this GKZ system is the affine Weyl group of the root system $A_\mu$. 
This is the mirror dual to the affine Weyl group which appears in the recent 
study of ${\rm Auteq} \,D(\widehat{\mathbf C^2/\mathbf Z_{\mu+1}})$ 
in \cite{IU}. 
}. 
\end{proposition}

\vskip0.3cm
The above result seems to be interesting from 
the viewpoint of the uniformization 
of hypergeometric series\cite{Yo}, since our GKZ system provides an infinite 
number of examples for which multi-valued hypergeometric series are 
uniformizable. However the result itself is not surprising, as 
we will show in the next section that our GKZ system has a close relation 
to K. Saito's differential equations for which the monodromy property 
is well-studied.

\vskip1cm
\section{ {\bf K.Saito's differential equations} }

In the deformation theory of a singularity, we have a notion of 
so-called {\it primitive form} which satisfies a set of differential 
equations, K. Saito's differential equations\cite{Sa}. 
Primitive form is 
also referred to as Gel'fand-Leray form\cite{AGV}. 
Here we briefly introduce the 
primitive form and K. Saito's differential equations, and then connect them 
to the `period integral' (\ref{eqn:Pi}) and the GKZ system 
(\ref{eqn:GKZdiff}). 

\vskip0.5cm
{\bf (4-1) K.Saito's system for a primitive form $\mathcal U_0(a)$.} 
Let us first note that, in the deformation of the singularity 
$U^2+V^2+W^{\mu+1}$, the polynomial 
equation $f_\Sigma(a,W)+U^2+V^2=0$ will be considered in $\mathbf C^3$ 
and the deformation parameters are set to $a_{\mu+2}=1, a_{\mu+1}=0$ 
by a coordinate change of $W$. 
Namely we take the defining equation of the form 
$f_\Sigma(a,W)+U^2+V^2=a_1+f_1(a,U,V,W)$ with  
$$
f_1(a,U,V,W)=a_2W+\cdots+a_{\mu}W^{\mu-1}+W^{\mu+1}+U^2+V^2 \;,
$$
and regard the parameters $a_1,a_2,\cdots,a_{\mu}$ as giving a deformation 
of the singularity $W^{\mu+1}+U^2+V^2=0 \subset \mathbf C^2$ at the origin. 
Since the parameter $a_1$ plays a distinguished role from the others, we set  
the local parameters $(a_2,\dots,a_{\mu})$ as a coordinate of 
$T:=\mathbf C^{\mu-1}$. The full parameters $(a_1,a_2,\cdots,a_{\mu})$ will 
be regarded as a coordinate of $S:=\mathbf C \times T$. 
We consider the product space $\mathfrak X:=\mathbf C^3 \times T$ 
with coordinate 
$(U,V,W,a_2,\cdots,a_{\mu})$.  Then we have a natural map 
$\varphi:\mathfrak X \rightarrow S$ by 
$$
(U,V,W,a_2,\cdots,a_{\mu}) \mapsto (-f_1(a,U,V,W),a_2,\cdots,a_{\mu})\;.
$$
This map plays important roles 
in describing the deformation of the singularity. 
Consider the sheaf $\Omega^p_{\mathfrak X/T}$ of germs of relative 
holomorphic $p$ forms for the natural projection 
$\pi: \mathfrak X =\mathbf C^3 \times T \rightarrow T$. 
We may consider the following sheaves on $S$,
$$
\mathcal H^{(0)}=\varphi_* \Omega^{3}_{\mathfrak X/T}/df_1 \wedge d(\varphi_*
\Omega^1_{\mathfrak X/T}) 
\;,\;\;
\mathcal H^{(-1)}=\varphi_* \Omega^{2}_{\mathfrak X/T}/
(df_1 \wedge \varphi_*\Omega^1_{\mathfrak X/T} + 
 d(\varphi_*\Omega^1_{\mathfrak X/T}) )\;.
$$
A primitive form $\zeta$ is an element in $H^0(S,\mathcal H^{(0)})$ 
satisfying certain conditions (, see \cite{Sa} for details). 
Instead of $\zeta$, hereafter, we consider its image 
$\mathcal U_0$ in $H^0(S,\mathcal H^{(-1)})$ 
under an isomorphisms $\mathcal H^{(0)} \cong \mathcal H^{(-1)}$ 
(see \cite{Sa}), 
which we may write explicitly as 
$$
\mathcal U_0(a)=Res_{\{a_1+f_1=0\}}\left( 
\frac{dU \wedge dV \wedge dW}{a_1+f_1(a,U,V,W)} \right) \;\;.
$$
In this from, the primitive form is also called as the Gel'fand-Leray form 
in the study of oscillating integrals (, see e.g. \cite{AGV}). We note that 
the similarity of $\mathcal U_0(a)$ to our `period integral' (\ref{eqn:Pi}), 
although, in (\ref{eqn:Pi}), we do not set $a_{\mu+2}=1, a_{\mu+1}=0$ but 
consider torus actions $(\mathbf C^*)^2$ instead.

K.Saito's system is defined as a set of differential equations 
satisfied by 
the primitive form $\mathcal U_0(a)=\mathcal U_0(a_1,\cdots,a_{\mu})$;  
\begin{eqnarray*}
&
P_{ij} \mathcal U_0(a)
&
=\left\{\frac{\pd^2\;}{\pd a_i \pd a_j} 
-\nabla_{\frac{\pd \;}{\pd a_i}}\frac{\pd\;}{\pd a_j} 
-(\frac{\pd\;}{\pd a_i}*\frac{\pd\;}{\pd a_j})\frac{\pd\;}{\pd a_1} 
\right\} \mathcal U_0(a)=0 \;\;, \\
&
Q(\frac{\pd\;}{\pd a_i})\mathcal U_0(a)
&
=
\left\{ w(\frac{\pd\;}{\pd a_i})\frac{\pd\;}{\pd a_1} 
- N(\frac{\pd\;}{\pd a_i}) + \frac{3}{2} \frac{\pd\;}{\pd a_i}\right\} 
\mathcal U_0(a) = 0 \;\;, 
\end{eqnarray*}
see \cite{Sa} for detailed definitions. This system is defined for 
a general setting in the deformation theory of singularities, and also known 
to define a holonomic system. A proof of the following proposition 
may be found in Appendix by Ambai in \cite{Oda1}, for example.  
Although, we restrict our attention to $A_{\mu}$ case, 
similar results are known also for other ($D$ or $E$) types of singularities.

\vskip0.5cm
\begin{proposition} \mylabel{thm:KsaitoSystem} Let $\beta_0(a),\cdots,
\beta_\mu(a)$ be the roots of $\psi_0(W):=a_1+a_2W+\cdots+a_{\mu}W^{\mu-1}
+W^{\mu+1}$ which satisfy $\beta_0(a)+\cdots+\beta_\mu(a)=0$. Then the 
space of solutions of K.Saito's system is generated by 
\begin{equation}
1 \;,\; 
\beta_0(a)-\beta_1(a) \;\;,\;\; \cdots \;,\;
\beta_{\mu-1}(a)-\beta_\mu(a) \;\;.\;\;
\end{equation}
The system has a regular singularity at the discriminant locus
$$
dis \psi_0(a)=\Pi_{1\leq i,j \leq \mu}(\beta_i(a)-\beta_j(a))^2 =0 \;\;,
$$
and the monodromy group about the discriminant coincides with 
the symmetric group $S_\mu$ of order $\mu$ acting as permutations 
among the roots $\beta_i(a)$. 
\end{proposition}
 
\vskip0.5cm

As we have noted in (3-1), $\mu$ vanishing cycles appears in the 
deformation of the singularity. 
Now it is easy to find that the above solutions 
$\alpha_i:=\beta_i(a)-\beta_{i-1}(a)$ represent the integrals 
$\int_C \mathcal U_0(a)$ over the corresponding 
vanishing cycles (, see Appendix A). Also we note that 
there is a residue pairing $I(\alpha_i,\alpha_j)$ among the 
solutions which reproduces the intersection 
pairing $\# L_i\cap L_j$ among the vanishing cycles \cite{Sa}.

\vskip0.5cm
{\bf (4-2)  GKZ system for a primitive form $\mathcal U(a)$.} 
As defined above, the primitive form $\mathcal U_0(a)$ is parametrized by 
$(a_1,\cdots,a_{\mu})\in S$ by setting $a_{\mu+1}=0, a_{\mu+2}=1$ in the 
defining equation $f_\Sigma(a,W)+U^2+V^2$. Instead of this, one may consider a 
natural torus actions $(\mathbf C^*)^2$ on $(a_1,\cdots, a_{\mu+2})$ by 
$a_i \mapsto \lambda a_i$ and $a_i \mapsto \lambda^{i-1} a_i$ 
$(\lambda \in \mathbf C^*)$. 
With this slight change of the parameters (`gauge'), we may connect K.Saito's 
system to a GKZ system. Let us define a period integral of the primitive form 
\begin{equation}
\int_C \mathcal U(a)=\int_C Res_{f_{\Sigma}(W)+U^2+V^2=0}\left(
\frac{dW \wedge dU \wedge dV}{f_\Sigma(a,W)+U^2+V^2} \right)  \;\;,
\mylabel{eqn:Pi2}
\end{equation}
where $C$ is a two cycle on $f_\Sigma(a,W)+U^2+V^2=0 \subset \mathbf C^3$ 
and $f_\Sigma(a,W)=a_1+a_2W+\cdots+a_{\mu+2}W^{\mu+1}$. The period integral 
above is similar to (\ref{eqn:Pi}), and thus satisfies 
a similar GKZ system to (\ref{eqn:GKZdiff}). 
The only difference appears in the 
scaling properties expressed by the linear operators $\mathcal Z_i$.

\vskip0.5cm
\begin{proposition} \mylabel{thm:GKZ2} 
1) The period integral (\ref{eqn:Pi2}) 
satisfies 
\begin{equation}
\Box_l \int_C \mathcal U(a)=0\;(l\in \mathcal L)\;\;,\;\;
\mathcal Z'_i \int_C \mathcal U(a)=0 \;(i=1,2) \;,
\mylabel{eqn:GKZdiff2}
\end{equation}
where the operators $\Box_l$ and the lattice $\mathcal L$ are the same as in 
(\ref{eqn:GKZdiff}), and $\mathcal Z_i' \,(i=1,2)$ are given by   
$$
\left(
\begin{matrix}
\mathcal Z'_1 \\
\mathcal Z'_2 \\  \end{matrix} \right)
=
\left(
\begin{matrix}
\theta_1+\theta_2+\cdots 
+\theta_{\mu+2} \\
\theta_2+2 \theta_3\cdots+(\mu+1)\theta_{\mu+2} +1\\
\end{matrix} \right)
$$
2) The system (\ref{eqn:GKZdiff2}) above is reducible of rank $\mu+1$ with 
its irreducible part of rank $\mu$. The $\mu$ independent solutions 
of the irreducible part are given by 
$$
\beta_0(a)-\beta_1(a),\cdots,\beta_{\mu-1}(a)-\beta_{\mu}(a) \;\;,
$$
where $\beta_i(a)$ are roots of 
$\psi(W)=a_1+a_2W+\cdots+a_{\mu+2}W^{\mu+1}=0$. 

\end{proposition}

\vskip0.5cm
Derivation of 1) above is straightforward, but it should be noted that  
K.Saito's system is replaced by a different but simpler GKZ system. 
Also, it is known that the numbers of independent solutions 
of a GKZ system is given by the volume of a relevant polytope\cite{GKZ1}, 
which in our case is $Vol(\Sigma)=\mu+1$. One may verify 
that the system is reducible observing the factorization of a differential 
operator in our GKZ system when expressing $\Box_l$ operators 
in the affine coordinates of $\mathcal M_{Sec(\Sigma)}$. 
After the factorization, the reduced GKZ system has $\mu$ independent 
solutions as claimed. 
(It has been observed in \cite{HKTY1} that the GKZ systems 
in mirror symmetry are often reducible in this way.)  One can derive 
the $\mu$ independent solutions in 2) by evaluating the period integral 
(\ref{eqn:Pi2}) over the vanishing cycles (, see Appendix A). Or more 
directly, one may verify that the roots 
$\beta_k$'s are solutions of ({\ref{eqn:GKZdiff2}) with one relation 
$\beta_0+\cdots+\beta_{\mu}=-\frac{a_{\mu+1}}{a_{\mu+2}}$ from the 
following property:

\vskip0.5cm
\begin{proposition} 
The root $\beta(a)\,(=\beta_0(a),\cdots,\beta_{\mu}(a))$ of $\psi(W)=0$ 
satisfies
\begin{equation}
\frac{\pd \beta}{\pd a_i}=-\frac{\beta^{i-1}}{\psi'(\beta)} \;\;,\;\;\;\;
\frac{\pd^2\beta}{\pd a_i \pd a_j}=\frac{1}{\psi'(\beta)}
\frac{d \;}{d x} \left( \frac{x^{i+j-2}}{\psi'(x)}\right)\Big|_{x=\beta} \;.
\mylabel{eqn:dbeta}
\end{equation}
\end{proposition}

\vskip0.3cm
\begin{proof} 
The first relation follows from the differentiation with respect to $a_i$ of 
$\psi(\beta(a))=a_1+a_2 \beta(a)+\cdot + a_{\mu+2} \beta(a)^{\mu+1}=0$. 
For the second relation, we note the following;
$$
\frac{\partial \psi'(\beta)}{\partial a_j}
=(j-1)\beta^{j-2}+\psi''(\beta)\frac{\partial \beta}{\partial a_j}
=(j-1)\beta^{j-2}-\frac{\psi''(\beta)}{\psi'(\beta)} \beta^{j-1} \;.
$$
Then the second relation follows by differentiating the 
first equation with respect to $a_i$. 
\end{proof}

\vskip0.5cm
Now a proof of Proposition \ref{thm:GKZ2} is straightforward 
by the relations above. Here, since the arguments are similar, 
we present the solutions of the GKZ system  
(\ref{eqn:GKZdiff}) for the local mirror symmetry, which is 
the same as the GKZ system (\ref{eqn:GKZdiff2}) for the primitive 
form except the scalings operators.

\vskip0.5cm
\begin{proposition} 
\mylabel{thm:GKZlogB}
1) The independent solutions of the GKZ system (\ref{eqn:GKZdiff}) 
are given by  
$$
1 \;\;,\;\; \log \beta_0(a)-\log \beta_1(a) \;\;,\;\; \cdots 
\;\;,\;\; \log \beta_{\mu-1}(a)-\log \beta_{\mu}(a) \;\;.
$$
2) These solutions, up to monodromy transformations, are 
related to the expansion $w(x;\frac{J}{2\pi i})=1+\sum_{k=1}^\mu w_{k}(a) J_k$ 
near the large complex structure by 
$$
2\pi i\, w_k(x)=\log \beta_{k-1}(a) - \log \beta_{k}(a)  \;\;(k=1,\cdots,\mu).
$$
\end{proposition}

\vskip0.3cm
\begin{proof} 
1) By the relations (\ref{eqn:dbeta}), it it straightforward 
to evaluate $\mathcal Z_1 \log \beta$ and $\mathcal Z_2 \log \beta$ as 
follows;
$$
\begin{aligned}
&
\mathcal Z_1 \log \beta 
= \frac{1}{\beta}\sum_{i=1}^{\mu+2} \theta_{a_i}\beta 
= -\frac{1}{\beta}\sum_{i=1}^{\mu+2} \frac{1}{\psi'(\beta)} a_i \beta^{i-1}
=-\frac{1}{\beta \psi'(\beta)}\psi(\beta)=0 \;\;,\\
&
\mathcal Z_2 \log \beta 
=\frac{1}{\beta}\sum_{i=1}^{\mu+2}(i-1)\theta_{a_i}\beta 
=-\frac{1}{\psi'(\beta)}\sum_{i=1}^{\mu+2}(i-1)a_i \beta^{i-2} 
=-1 \;.
\end{aligned}
$$
Therefore the differences $\log \beta_k -\log \beta_{k-1}$ $(k=1,..,\mu)$ 
are annihilated by  the linear operators $\mathcal Z_1, \mathcal Z_2$. 
As for the operators $\Box_l \; (l \in \mathcal L)$, we first note  
\begin{equation}
\frac{\partial^2 \; }{\partial a_i \partial a_j} \log \beta 
= 
\frac{\partial^2 \; }{\partial a_k \partial a_l} \log \beta  
\;\; \text{ if } \;  i+j=k+l \;, 
\mylabel{eqn:daij}
\end{equation}
which follows from 
$$
\frac{\partial^2 \; }{\partial a_i \partial a_j} \log \beta 
=-\frac{1}{\beta^2} \frac{\beta^{i+j-2}}{(\psi'(\beta))^2} 
 +\frac{1}{\beta}\frac{1}{\psi'(\beta)} 
\frac{d \;}{d x} \left( \frac{x^{i+j-2}}{\psi'(x)}\right)\Big|_{x=\beta} \;.
$$
Since the entries of $l=(l_1,l_2,...,l_{\mu+2})$ satisfies $\sum_i l_i =0$ 
and $\sum_i (i-1)l_i =0$, we have $\sum_i i \, l_i^+ = \sum_i i \, l_i^-$ 
for $l_i^\pm:=\frac{1}{2}(l_i\pm |l_i|)$. Now, by repeated use 
of (\ref{eqn:daij}), one may deduce 
$$
\prod_i  
\left(\frac{\partial \; }{\partial a_i}\right)^{l_i^+} \log \beta 
=
\prod_i  
\left(\frac{\partial \; }{\partial a_i}\right)^{l_i^-} \log \beta \;,  
$$
which means $\Box_l \log \beta =0$. 
Since the linear independence of $\log \beta_k -\log \beta_{k-1}$ 
$(k=1,..,\mu)$ is clear, we obtain the claim including the trivial 
solution $1$. 

2) 
By definition, we have $ 2\pi i \, w_k(x) \sim \log x_k$, 
where $x_k=\frac{a_k a_{k+2}}{a_{k+1}^2}$. Since, by extending 
the classical arguments due to Frobenius, it is straightforward to show 
that $w_k(x)$ are solutions of the GKZ hypergeometric system 
(\ref{eqn:GKZdiff}), we only need to connect these solutions to 
those in terms of roots $\beta_k$. For this, let us assume that the roots are 
given by $\beta_k=-\frac{1}{\lambda_k}$ and also the following relation, 
$$
\psi(W)=a_1+a_2 W + \cdots + a_{\mu+2}W^{\mu+1} = 
\prod_k (\lambda_k W + 1) \;\;.
$$
Assume also that $|\lambda_{\mu}|<<|\lambda_{\mu-1}|<<\cdots<<|\lambda_0|$. 
Then from the relations between roots and the coefficients, it is easy 
to see 
$$
\begin{aligned}
\log a_1 =0 \;,\; 
&\log a_2 = \log \lambda_0 + A_2(\lambda) \;,\; 
\log a_3 = \log (\lambda_0\lambda_1) + A_3(\lambda) \;,\; 
\cdots \;\;, \\
&\log a_{\mu+2} = \log (\lambda_0\lambda_1\cdots \lambda_{\mu}) + 
A_{\mu+2}(\lambda) \;,\; \\
\end{aligned}
$$
where $A_k(\lambda) \in \mathbf C \{ 
\frac{\lambda_1}{\lambda_0}, 
\frac{\lambda_2}{\lambda_1}, 
\cdots, \frac{\lambda_\mu}{\lambda_{\mu-1}} \}$. Now we have 
$$
\log x_k = \log \frac{a_k a_{k+2}}{a_{k+1}^2} = 
\log \lambda_k - \log \lambda_{k-1} + B_k(\lambda) \;, 
$$
with $B_k(\lambda):=A_k(\lambda)+A_{k+2}(\lambda)-2 A_{k+1}(\lambda)$, 
which entails the claimed relation 
since $2\pi i w_k(x) \sim \log x_k$. 
(Note that if we invert the series relations 
$x_k=\frac{\lambda_k}{\lambda_{k-1}}e^{B_k(\lambda)}$ in 
$C\{ \frac{\lambda_1}{\lambda_0}, 
\frac{\lambda_2}{\lambda_1}, 
\cdots, \frac{\lambda_\mu}{\lambda_{\mu-1}} \}$, then we obtain 
$\frac{\lambda_k}{\lambda_{k-1}}$'s in terms of $x_k$'s, and thus recover the 
series expansion of $2\pi i w_k(x)$ by 
$2\pi i w_k(x)=\log \lambda_k - \log \lambda_{k-1}=
\log x_k - B_k(\lambda)$.) 
\end{proof}

\vskip0.8cm
The above form of the solutions may be connected to the piecewise linear 
functions on the fan $\Sigma$ \cite{Ba3}. 
Since we have established a relation of the GKZ system (\ref{eqn:GKZdiff}) 
of $\widehat{\mathbf C^2/\mathbf Z_{\mu+1}}$ 
to K.Saito's system of the primitive form, for which integral monodromy 
property is well studied  (see Proposition \ref{thm:KsaitoSystem}), 
it is clear that we have the uniformization property of the mirror map 
stated in Proposition \ref{thm:Uniform}.

\vskip1cm
\section{ {\bf Central charge formula and  G-Hilb} }

In the last two sections, we have looked at the local mirror symmetry of 
$\widehat{\mathbf C^2/\mathbf Z_{\mu+1}}$ paying our attentions to the 
monodromy property of the associated GKZ system. Here we come back to 
our claim for the central charge formula (\ref{eqn:chargeWxJ}) in this case.

\vskip0.3cm
Let us first recall that the non-compact Calabi-Yau manifold 
$X=\widehat{\mathbf C^2/\mathbf Z_{\mu+1}}$ is given as the Hilbert scheme 
of points on $\mathbf C^2$, G-Hilb $\mathbf C^2$. Here G-Hilb $\mathbf C^n$ 
is defined for a finite subgroup $G \subset SL(n,\mathbf C)$ and consists 
of zero dimensional subschemes $Z$ in $\mathbf C^n$ of length equal $|G|$ 
such that $G$ acts on $Z$ and $H^0(\mathcal O_Z)$ as the regular 
representation of $G$. The following results 
are due to Gonzalez-Sprinberg and Verdier\cite{GV} for $n=2$ and their 
generalizations to $n=3$ have been done in \cite{Na}\cite{IN}\cite{BKR}. 
Here, we summarize the relevant results for our cases where $G \subset 
SL(n,\mathbf C)$ is a finite abelian subgroup:

\vskip0.5cm
\begin{list}{}{
\setlength{\leftmargin}{12pt}
\setlength{\labelwidth}{12pt}
}
\item[(i)] The K-group $K(X)$ of algebraic vector bundles on $X$ 
are generated by 
the so-called tautological bundles $\mathcal F_0,\mathcal F_1,\cdots,
\mathcal F_{d}$ $(d=|G|-1)$, where the subscripts refer to the 
one-dimensional representations of $G$ with $0$ for the trivial 
representation. 

\item[(ii)] Let $K^c(X)$ be the K-group of the complexes of 
algebraic vector bundles which are exact off $\pi^{-1}(0)$ where 
$\pi:X \rightarrow \mathbf C^n/G$. Then there exists a complete pairing 
$K^c(X) \times K(X) \rightarrow \mathbf Z$, and the dual 
basis $S_0,S_1,\cdots,S_d$ of $K^c(X)$ satisfying
\begin{equation}
\langle ch(S_i),ch(\mathcal F_j) \rangle :=\int_X ch(S_i)ch(\mathcal F_j) 
Todd_X = \delta_{ij} \;\;.
\mylabel{eqn:StoF}
\end{equation}

\item[(iii)] The dual basis $S_k (k=0,\cdots,d)$ defines a symmetric form 
for $n=2$ (, a symplectic form for $n=3$,) on $K^c(X)$ by
\begin{equation}
\chi(S_i,S_j):=\int_X ch(S_i^\vee)ch(S_j)Todd_X 
\;\;(0 \leq i,j \leq d).
\mylabel{eqn:sympS}
\end{equation}
In two dimensional case with $G=\mathbf Z_{\mu+1}$, $S_k$'s for 
$k\not=0$ have a simple forms. They are given by 
$S_k=\mathcal O_{C_k}(-1)$ with $C_k \cong \mathbf P^1$ being the 
exceptional curves. Including $S_0$, they satisfy 
$\chi(S_i,S_j)=-\hat{\mathcal C}_{ij} 
\;\;(0\leq i,j \leq d)$ where $\hat{\mathcal C}_{ij}$ is the extended 
Cartan matrix of the root system of type $A_\mu$. 
\end{list}

\vskip0.5cm

Now with these properties about the geometry of $X=\widehat{\mathbf C^2/
\mathbf Z_{\mu+1}}$, let us recall our Conjecture \ref{thm:conj}; 
$$
w(x_1,\cdots,x_r;\frac{J_1}{2\pi i},\cdots,\frac{J_r}{2\pi i})=
\sum_{i,j} \int_{mir(E_i)} \Omega(Y_x) \; \chi^{ij} \; ch(E_j^\vee) \;.
$$
One should note that this is a conjecture for $X$ a compact Calabi-Yau 
manifold. It is rather clear how to modify this 
relation for our non-compact $X$ if we notice that the expression 
$\sum_j \chi^{ij}ch(E_j^\vee)$ satisfies $\int_X ch(E_k) 
\sum_j \chi^{ij}ch(E_j^\vee) Todd_X = \delta^{i}_{\;k}$, i.e., 
provides a dual base to $ch(E_k)$. From this, 
we claim that the central charge formula for 
$X=\widehat{\mathbf C^2/\mathbf Z_{\mu+1}}$ is given by 
\begin{equation}
w(x_1,\cdots,x_\mu;\frac{J_1}{2\pi i},\cdots,\frac{J_\mu}{2\pi i})=
\sum_{k} \int_{mir(S_k)} \Omega(Y_x) \; ch(\mathcal F_k) \;,
\mylabel{eqn:WxJlocal}
\end{equation}
where the holomorphic two form $\Omega(Y_x)$ should be understood 
as a holomorphic two-form associated to 
the integrand of (\ref{eqn:Pi}), i.e., 
$$
\Omega(Y_x)=\frac{i}{2\pi^2} Res\left( 
\frac{1}{f(W)+U^2+V^2}dU dV \frac{dW}{W} \right) \;\;.
$$ 
We will state the corresponding claim in 
Conjecture 6.3 for three dimensional cases.

\vskip0.3cm

Now, since the tautological bundles satisfy  
$$
\int_{C_j} c_1(\mathcal F_k)=\delta_{jk} \;\;(1 \leq j,k \leq \mu),
$$
for the exceptional curves $C_j \,(1\leq j \leq \mu)$, 
we have $c_1(\mathcal F_k)=J_k \;(k\geq 1)$ for our generators 
$J_k$ of $H^2(X,\mathbf Z)$. Therefore we have 
$$
w(x;\frac{J}{2\pi i})=1+\sum_k w_k(x) J_k
=(1-\sum_{k=1}^\mu w_k(x) )ch(\mathcal F_0)
 + \sum_{k=1}^\mu w_k(x) ch(\mathcal F_k) \;,
$$
where we use $ch(\mathcal F_k)=1+c_1(\mathcal F_k)=1+J_k$. 
From this, one may read  the central charge of 
the sheaf $S_k=\mathcal O_{C_k}(-1) \, (k\not=0)$ as
\begin{equation}
Z_t(S_k)=\int_{mir(S_k)} \Omega(Y_x) = w_k(x)=
\frac{1}{2\pi i}\big( \log \beta_{k-1}(a)-\log \beta_{k}(a) \big) \;\;. 
\mylabel{eqn:ZSk}
\end{equation}
In Appendix A, we obtain the above central charge as a period integral 
over a vanishing cycle. Therefore, we have obtained a refinement of 
the mirror symmetry sketched in (3-1) that the mirror image  
$mir(S_k)$ of the sheaf $S_k=\mathcal O_{C_k}(-1)$ is a  
vanishing cycle whose period is given by $Z_t(S_k)$ above. 

\vskip0.3cm
Similar relation holds also for $S_0$ with its central charge 
$1-\sum_k w_k(x)$. However, since $S_0$ is not a sheaf but a complex of 
sheaves (see \cite{IN}), 
we replace it with the skyscraper sheaf $\mathcal O_p$ 
supported at a point $p$. 
Then the sheaves $\mathcal O_p, \mathcal O_{C_k}(-1) \; 
(k=1,\cdots,\mu)$ form another basis of $K^c(X)$ with relations 
$$
\mathcal O_p=S_0+S_1+\cdots+S_\mu\;,\;\; 
\mathcal O_{C_k}(-1)=S_k \;(k=1,\cdots,\mu) \;,
$$
where we use $\langle \mathcal O_p, \mathcal F_j \rangle=1 \;
(j=0,\cdots,\mu)$ to derive the first equality. 
In (6-2), this basis of $K^c(X)$ will be generalized as a 
{\it symplectic D-brane basis} to the three-dimensional cases. 
Now, for the sheaf $\mathcal O_p$, we have 
$$
Z_t(\mathcal O_p)=\int_{mir(\mathcal O_p)}\Omega(Y_x) \equiv 1 \;\;.
$$
This is identified with the period integral over a cycle $T_0$ 
whose topology is $S^1 \times S^1$, see Appendix A.

\vskip0.5cm
Finally, we note that the sheaves $S_k$ $(k\not=0)$ are spherical 
and thus define 
self-equivalences of $D^b(Coh(X))$, the Seidel-Thomas twists summarized in 
section 2. In [ST, Proposition 3.19], it is shown that these spherical 
objects, which form the so-called $(A_\mu)$-configuration, 
generate a weak braid group action on $D^b(Coh(X))$. 
This braid group action should be mirror to the corresponding 
Dehn twists (Picard-Lefschetz transformations) in the symplectic side. 
Our formula (\ref{eqn:ZSk}) explicitly shows this mirror correspondence as 
the linear transformations on the central charges.

\vskip1cm
\vfill\eject
\section{ {\bf Local mirror symmetry II --- $X=\widehat{\mathbf C^3/G}$ }}

In the two dimensional cases, the local mirror symmetry is rather clear 
from the hyperk\"ahler rotation as summarized in (3-1), and our results 
in the previous sections are consistent with those arguments from the 
hyperk\"ahler rotation. In this section, we will argue that our central 
charge formula for a crepant resolution of the three dimensional 
singularity $\mathbf C^3/G$ with a finite abelian group $G \subset 
SL(3,\mathbf C)$. The crepant resolution we use is the so-called 
$G$-Hilb $\mathbf C^3$ whose definition has been given in the 
previous section. Examples of GKZ hypergeometric system in this 
situation have been studied also in \cite{dOFS}.

\vskip0.5cm
{\bf (6-1) Period integrals and GKZ systems.} 
As above, let us consider a non-compact Calabi-Yau manifold 
$X=G$-Hilb $\mathbf C^3$ with a finite, abelian group 
$G \subset SL(3,\mathbf C)$. For $G$ being abelian, $X$ is given as a 
toric resolution of a singularity $\mathbf C^3/G$ as follows:
Let us first denote by $M_G$ the lattice corresponding to the invariants 
$\mathbf C[x^{\pm},y^{\pm},z^{\pm}]^G$, and its dual lattice by $N_G$. 
Then the invariant monomials in $\mathbf C[x,y,z]^G$ defines a cone 
in $M_G\otimes \mathbf R$, with its dual cone 
$\Sigma_G^0$ in $N_G\otimes \mathbf R$. The toric (Calabi-Yau) resolution 
which corresponds to $G$-Hilb $\mathbf C^3$ is given by a fan obtained by 
subdividing the cone $\Sigma_G^0$, which we will denote 
by $\Sigma_G$ hereafter. (See \cite{Re} for detailed construction of 
the fan $\Sigma_G$.) 
We will denote by $\nu_1, \nu_2, \cdots, \nu_{r+3}$ the integral generators 
of the one-dimensional cones in the fan $\Sigma_G$. Note that, since 
the resolution is crepant, these generators satisfy 
$$
( m, \nu_k ) = 1  \;\;(k=1,\cdots,r+3) \;,
$$
with some $m \in M_G$, where $(\;,\;): M_G \times N_G \rightarrow \mathbf Z$ 
is the dual pairing.  

To describe the mirror configuration of $X$, let us fix an isomorphism 
$\varphi: N_G \simeq \mathbf Z^3$ satisfying $\varphi(\nu_k)=(1,\bar \nu_k)$. 
Consider the defining equation 
\begin{equation}
F(a;U,V,W_1,W_2)=U^2+V^2+\sum_{k} a_k W^{\bar\nu_k} \;\;,
\mylabel{eqn:mirrorF}
\end{equation}
then the mirror configuration of $X$ claimed in \cite{CKYZ}\cite{HIV} is   
the hypersurface $Y=(F(a;U,V,W_1,W_2)=0) \subset \mathbf C^2 \times 
(\mathbf C^*)^2$. Note that $Y$ does not depend on the choice of the 
isomorphism $\varphi: N_G \simrightarrow \mathbf Z^3$ 
with the described property, 
since $W_1, W_2$ are considered in $\mathbf C^*$ and negative powers of them 
are allowed.  One important criterion for the mirror configuration $Y$ 
is the fact that we can extract the right Gromov-Witten invariants 
for $X$ from the ``period integrals'' of $Y$ (cf. \cite{CKYZ}).

\vskip0.5cm
\begin{definition} For the mirror configuration $Y$ given above, we 
define the period integral of a cycle $L \in H_3(Y,\mathbf Z)$ by 
\begin{equation}
\Pi_L(a) = \frac{1}{4\pi^3} 
\int_L Res_{F=0} \left( \frac{1}{U^2+V^2+\sum_{k} a_k W^{\bar\nu_k}}
 dU dV \frac{dW_1}{W_1} \frac{d W_2}{W_2} \right) \;,
\mylabel{eqn:3-dimP}
\end{equation}
where the residue is taken around $F(a,U,V,W)=0$ in $\mathbf C^2 \times 
(\mathbf C^*)^2$. 
\end{definition}

\vskip0.5cm
In (6-3), we will describe how one can determine the Gromov-Witten 
invariants from our cohomology-valued hypergeometric series.
Here, as in the two dimensional cases, we see that the following GKZ system 
is satisfied by the period integral:

\vskip0.5cm
\begin{proposition} 
\mylabel{thm:propGKZ3}
Let $\mathcal L$ be the lattice of relations defined by 
$$
\mathcal L=\{ (l_1,\cdots,l_{r+3}) \in \mathbf Z^{r+3} \;|\; 
   \sum_k l_k \nu_k = \vec 0 \;\}.
$$
Then the period integral (\ref{eqn:3-dimP}) satisfies the following 
set of differential equations:
\begin{equation}
\Box_l \Pi_L(a)=0\;(l\in \mathcal L)\;\;,\;\;
\mathcal Z_i \Pi_L(a) =0 \;(i=1,2,3) \;,
\mylabel{eqn:3dimGKZ}
\end{equation}
where 
$$
\Box_l=\left(\frac{\pd\;}{\pd a}\right)^{l^+}-
\left(\frac{\pd\;}{\pd a}\right)^{l^-}\;\;,\;\;
\left(
\begin{matrix}
\mathcal Z_1 \\
\mathcal Z_2 \\  
\mathcal Z_3 \\  
\end{matrix} \right)
=
\left(
\begin{matrix}
\theta_1+\theta_2+\cdots 
+\theta_{r+3} \\
\sum_k \bar\nu_{k,2} \theta_k \\
\sum_k \bar\nu_{k,3} \theta_k \\
\end{matrix} \right)
$$
with $l=l^+-l^-$ $(l^\pm_i:=\frac{1}{2}(l_i \pm |l_i|) \,)$ and  
$\theta_k=a_k\frac{\pd\;}{\pd a_k}$.  $\bar\nu_{k,i}$ 
represents the $i$-th component of the vector $\varphi(\nu_k)=(1,\bar 
\nu_k)$ with the isomorphism $\varphi: N_G \simrightarrow \mathbf Z^3$ 
fixed above.
\end{proposition}

\vskip0.5cm
\noindent
{\bf Remark.} 
As we have shown in section 4, by a slight modification of the period 
integral, we can connect (\ref{eqn:3-dimP}) to the K. Saito's 
primitive form of a three dimensional singularity. The primitive form 
satisfies the same GKZ hypergeometric system but with a different 
scaling property, i.e., $\mathcal Z_i (i=2,3)$ being replaced by 
$\mathcal Z_i':=\mathcal Z_i+1 (i=2,3)$. 
One may observe for several examples that, 
when the singularity of 
$\mathbf C^3/G$ is isolated, the resulting GKZ system for 
the primitive form is reducible and its irreducible part  
shows a slightly different degeneration at `infinity', i.e. at 
the large complex structure limit (, see Appendix B). 
\hfill \qed

\vskip0.5cm
{\bf (6-2) Central charge formula and symplectic forms.} 
The cohomology-valued hypergeometric series for the GKZ system 
(\ref{eqn:3dimGKZ}) are defined by constructing the secondary 
fan $Sec(\Sigma_G)$. 
As described for an example in (3-3) and Fig.1, the secondary fan 
is constructed combinatorially from the secondary polytope in 
(a translate of) $\mathcal L_{\mathbf R}$. 
We refer \cite{HLY},\cite{Hos},\cite{Sti} and references therein for details 
of the construction in our context of mirror symmetry. 

Let us denote by $l^{(1)}, \cdots, l^{(r)}$ the integral generators 
of $\mathcal L$ which represents the large complex structure limit. 
We denote 
the affine coordinate by $x_k=a^{l^{(k)}} \; (k=1,\cdots,r)$. Then, 
we have cohomology-valued hypergeometric series 
\begin{equation}
w(x_1,\cdot\cdot,x_r; \frac{J_1}{2\pi i},\cdot\cdot,\frac{J_r}{2\pi i})
=\sum_{m \in \mathbf Z_{\geq 0}^s}  c(m+\rho)
x^{m+\rho} 
\vert_{\rho=\frac{J}{2\pi i}},  
\mylabel{eqn:covGKZ}
\end{equation}
where $c(m):=c(m_1,\cdots,m_r)$ is defined by 
$$
c(m_1,\cdots,m_r):=
\frac{1}{\prod_{i=1,\dots,r+3}\Gamma(1+\sum_k m_k l^{(k)}_i)} \;\;,
$$
and $J_1,\cdots,J_r \in H^2(X,\mathbf Z)$ are dual (semi-ample) generators 
to $l^{(1)},\cdots,l^{(r)}$ (, see [Sect.1, Hos] for a quick review of 
the construction). It is found in \cite{HKTY2} for $X$ compact 
Calabi-Yau hypersurfaces (see also \cite{Sti}), and used in \cite{CKYZ} for 
non-compact cases that the above cohomology-valued hypergeometric series 
generate the solutions of the GKZ system (\ref{eqn:3dimGKZ}) when expanded 
in $H^{even}(X,\mathbf Q)$. If we recall the integral structure of 
$K(X)$ and also the natural symplectic structure 
in $K^c(X)$ which are summarized in section 4, 
we come to the following conjecture:

\vskip0.5cm
\begin{conjecture} 
\mylabel{thm:conj3}
1)  Consider the expansion of the cohomology-valued hypergeometric 
series (\ref{eqn:covGKZ}) with respect to the basis $ch(\mathcal F_0), 
\cdots, ch(\mathcal F_d)$ of $H^{even}(X,\mathbf Q)$;
$
w(x;\frac{J}{2\pi i})=\sum_k f_k(x) ch(\mathcal F_k).
$
Then the coefficient hypergeometric series $f_k(x)$ may be identified 
with the period integrals over the cycles $mir(S_k)$, i.e., we have 
\begin{equation}
w(x_1,\cdots,x_r;\frac{J_1}{2\pi i},\cdots,\frac{J_r}{2\pi i})=
\sum_{k} \int_{mir(S_k)} \Omega(Y_x) \; ch(\mathcal F_k) \;,
\mylabel{eqn:WxJlocal3}
\end{equation}
where $mir: K^c(X) \simrightarrow H_3(Y,\mathbf Z)$ is the (homological) 
mirror map and $\Omega(Y_x)=\frac{1}{4\pi^3} Res_{F=0}
(\frac{1}{F(a,U,V,W)}
dUdV\frac{dW_1}{W_1}\frac{dW_2}{W_2})$. 

\noindent
2) The monodromy of the coefficient hypergeometric series $f_k(x) 
(=\int_{mir(S_k)} \Omega(Y_x))$ is integral and symplectic with 
respect to the symplectic form; 
$$
\chi(S_i,S_j)=\int_X ch(S_i^\vee)ch(S_j) Todd_X \;\; 
(0 \leq i,j \leq d=|G|-1).
$$

\noindent
3) Using the cohomology-valued hypergeometric series, the central 
charge of $F \in K^c(X)$ is given by $Z_t(F)=\int_X ch(F) 
w(x;\frac{J}{2\pi i}) Todd_X$. 

\end{conjecture}

\vskip0.5cm

In section 4 and section 5, we have verified explicitly 
the corresponding conjecture for 
$X=\widehat{\mathbf C^2/\mathbf Z_{\mu+1}}$ by connecting our 
period integral to that of the primitive form in singularity theory. 
For three dimensions, however, little is known about the explicit 
form of the period integral of the primitive form. 
In the following sections, we will test 
our conjecture for the cases, $G=\mathbf Z_3, \mathbf Z_5$. 
In particular, for the former case, we will evaluate the period integrals 
of the primitive form, and connect them to the periods in mirror symmetry 
(, see Appendix A). In this case, we will be able to see 
the mirror correspondence $mir: K^c(X) \simrightarrow H_3(Y,\mathbf Z)$ 
from our Conjecture \ref{thm:conj3}, which generalizes the correspondence 
between the spherical objects and the vanishing cycles 
presented in section 5. 

As another aspect of our conjecture, we will define the so-called 
prepotential which gives predictions of Gromov-Witten invariants 
on the geometry $X$. When $X$ is compact, the existence of the 
prepotential (special K\"ahler geometry) is ensured by a canonical 
symplectic basis of $H_3(Y,\mathbf Z)$ and also by the property 
called Griffiths transversality \cite{St}. Correspondingly, in our 
non-compact cases, we have the symplectic structure on 
$H_3(Y,\mathbf Z)$ given by $(\chi(S_i,S_j))$ under the 
homological mirror map, $mir: K^c(X) \simrightarrow H_3(Y,\mathbf Z)$. 
We will find that our cohomology-valued hypergeometric series 
combined with the symplectic structure suffices to present 
a closed formula of the prepotential.

\vskip0.5cm
{\bf (6-3) Prepotential and a symplectic $D$-brane basis.} 
In string theory, the so-called D-branes 
play important roles for duality symmetries such as mirror symmetry. 
In our non-compact Calabi-Yau manifolds, roughly speaking, the D-branes 
are elements in $K^c(X)$ or more precisely elements in the  
category $D^b_c(Coh(X))$ of compact support. In what follows, 
we will work on $K^c(X)$( or $H_c^{even}(X,\mathbf Q)$ taking the Chern 
character). 
In our case, it is easy to see that there are natural D-branes 
like $\mathcal O_p$, the skyscraper sheaf, $\mathcal O_C$, 
the torsion sheaf supported on a curve $C$, and 
$\mathcal O_D$, that supported on an exceptional divisor $D$. Depending on the 
dimensions of the support, these sheaves are called $D0$-, $D2$-, $D4$-branes, 
respectively, see \cite{Do} and references therein. We will connect the bases 
$S_i$'s to these $D$-branes.

\vskip0.3cm
\noindent
(1) {\it Symplectic D-brane basis and its dual}:  
First, let us recall that we have 
introduced the (semi-ample) generators $J_1, \cdots, J_r$ of 
$H^2(X,\mathbf Z)$ when writing down our cohomology-valued 
hypergeometric series. Assume that these generators measure the volume 
of the respective curves $C_1,\cdots, C_r$, i.e., 
$$
\int_{C_i} J_k = \delta_{ik} \;\;.
$$
Assume also that the exceptional divisors $D_1,\cdots,D_s$ provide a 
basis of $H_4(X,\mathbf Z)$. Then in such cases, we seek a basis 
of $K^c(X)$ of the form, 
\begin{equation}
\{ \mathcal B_I \}:=
\{\; \mathcal O_p \,;\, \mathcal O_{C_1}(-J_1), \cdots,  
\mathcal O_{C_r}(-J_r) \,;\,  
\mathcal O_{D_1}(\mathcal L_1), \cdots,  
\mathcal O_{D_s}(\mathcal L_s) \;\;\}
\mylabel{eqn:Dbasis}
\end{equation}
where $\mathcal O_{C}(-J)=\mathcal O_{C}\otimes \mathcal O_X(-J)$ and 
$\mathcal O_{D_k}(\mathcal L_k):=
\mathcal O_{D}\otimes \mathcal L_k$ with $\mathcal L_k=
\mathcal O_X(-a_{k,1}J_1-\cdots -a_{k,r}J_r)$. 
Here we fix the integers $a_{k,1},...,a_{k,r}$ requiring   
\begin{equation}
\chi(\mathcal O_{D_j}(\mathcal L_j),\mathcal O_{D_k}(\mathcal L_k))=0
\mylabel{eqn:vanishing}
\end{equation}
for $j,k=1,\cdots,s$. 
The integers $a_{k,1},...,a_{k,r}$ satisfying the above requirement 
are not unique in general. 
However the reason of our requirement (\ref{eqn:vanishing}) 
will become clear soon when we present a closed formula of the prepotential. 
Hereafter, we will call a basis $\{ \mathcal B_I \}$ satisfying the  
requirement as a {\it symplectic $D$-brane basis} of $K^c(X)$. 

Now let us consider the dual basis of $K(X)$  
to the basis $\{\mathcal B_I \}$ (\ref{eqn:Dbasis}) under the 
paring (\ref{eqn:StoF}). 
Taking $ch: K(X) \rightarrow H^{even}(X,\mathbf Q)$, we construct 
(charges) $\{ Q_J \} :=\{Q_0;Q_2^1,..,Q_2^r;Q_4^1,..,Q_4^s\}$ 
which satisfies
$$
\int_X ch(\mathcal B_I)\,  Q_J \, Todd_X= \delta_{IJ} \;\;.
$$
We note that, due to the support property of the D-brane basis 
(\ref{eqn:Dbasis}) and the above relation, 
the base $Q_0$ starts from the degree zero 
element $1$ to higher degree terms. Similarly $Q_2^a$ starts from $J_a$ 
to higher degree terms, and $Q_4^b$ is written by a linear combination 
of the degree four elements of $H^{even}(X,\mathbf Q)$. 
The set of charges $\{ Q_J \}$ provides a basis 
of $H^{even}(X,\mathbf Q)$.

\vskip0.5cm
\noindent
(2) {\it Expansion of $w(x;\frac{J}{2\pi i})$ and the prepotential}: 
Now we expand our cohomology-valued hypergeometric series using 
the basis $\{ Q_J \}$ of $H^{even}(X,\mathbf Q)$; 
$$
w(x;\frac{J}{2\pi i})=
w_{0}(x) Q_0 + 
\sum_{a=1}^r w_{a}^{(1)}(x) Q_2^a + 
\sum_{b=1}^s w_{b}^{(2)}(x) Q_4^b \;\;.
$$
Then our conjecture says that the coefficient hypergeometric series 
$\{ w_0(x), w_a^{(1)}(x),$ $w_b^{(2)}(x) \}$ represents the central charges 
of the sheaves $\{ \mathcal B_I \}$ in (\ref{eqn:Dbasis}) and whose monodromy 
matrices are integral and symplectic with respect to the symplectic form 
\begin{equation}
( \chi(\mathcal B_I, \mathcal B_J) ) = 
\left( 
\begin{matrix}
0  & 0  & 0        \\
0  & 0  & C_{ab}  \\
0  &-C_{ab} &0  \\
\end{matrix} 
\right),
\mylabel{eqn:sympC}
\end{equation}
where we set $C_{ab}=\chi(\mathcal O_{C_a}(-J_a), \mathcal O_{D_b}
(\mathcal L_b))$ and 
used the fact $\chi(\mathcal O_p,\mathcal B_I)=0$ for sheaves 
$\mathcal B_I$ of compact support.  The above form  
should be contrasted to the canonical symplectic form on $H_3(Y,\mathbf Z)$ 
for the compact (mirror) Calabi-Yau manifolds, by which we define the 
special K\"ahler geometry on the moduli spaces(, see e.g. \cite{St}). 
In fact, we may define the 
prepotential for local mirror symmetry given in \cite{CKYZ} in terms of the 
symplectic form (\ref{eqn:sympC}).

For simplicity, let us assume 
$\dim H_c^{2}(X,\mathbf Q)=\dim H_c^{4}(X,\mathbf Q)$ (, i.e. $r=s$), and 
thus $(C_{ab})$ is a square matrix. Also we assume $\det (C_{ab})\not=0$. 
In this case, inverting the matrix $(C_{ab})$, we may define  
the `symplectic duals' of 
$\mathcal O_{C_1}(-J_1),\cdots,\mathcal O_{C_r}(-J_r)$ by 
$\sum_b C^{ab} \mathcal O_{D_b}(\mathcal L_b)$ $(a=1,\cdots,r)$ where 
$(C^{ab}):=(C_{ab})^{-1}$. We should note that these symplectic duals 
are not in $K^c(X)$ in general, but are elements in $K^c(X)\otimes \mathbf Q$. 
Correspondingly we consider the symplectic duals of the mirror cycles 
$mir(\mathcal O_{C_a}(-J_a))$ by $\sum_b C^{ab} mir(\mathcal O_{D_b}
(\mathcal L_b))$. From the period integrals of these cycles, we define 
the prepotential $F(t)$ by the special K\"ahler geometry relation;
\begin{equation}
(\, 1 \,,\, t_a  \,,\, \frac{\partial F}{\partial t_a}\,) = 
(\, 1 \,,\, w^{(1)}_a(x) \,,\, \sum_b C^{ab} w^{(2)}_b(x) \,)\;,
\mylabel{eqn:specialK}
\end{equation}
where we use the fact $w^{(0)}(x)=Z(\mathcal O_p)\equiv 1$ 
for local mirror symmetry of $X=\widehat{\mathbf C^3/G}$ (cf. section 5). 
Integrating this special K\"ahler geometry relation, we obtain 
the prepotential, which gives the right predictions for Gromov-Witten 
invariants(, see \cite{CKYZ} for examples). 

For general $G$, the matrix $(C_{ab})$ may not be square. Even if it is square 
it might be singular. 
In such cases, some of the sheaves $\mathcal O_{C_a}(-J_a)$ 
do not have its symplectic duals. However the special K\"ahler relation 
(\ref{eqn:specialK}) still makes sense under suitable modifications.

\vskip0.5cm
{\bf (6-4) Examples.} 
Here we will present two examples to verify Conjecture \ref{thm:conj3}. 
Some detailed physical analysis of D-branes and $\Pi$ stability on the first 
example may be found in \cite{DFR} (, see also \cite{Ma}\cite{Do} and 
references therein).  The hypergeometric series (with a different 
definition) of the second example are also studied in \cite{dOFS} (see also 
\cite{MR}). 

\vskip0.3cm
\noindent
{\it Example (1)}: 
As the first example, we consider the case $X=\widehat{\mathbf C^3/
\mathbf Z_3}$ with the group action specified by the weights 
$\frac{1}{3}(1,1,1)$. The toric resolution is 
the same as the $G$-Hilb and may be identified with the total 
space of the canonical bundle $K_{\mathbf P^2}$. 
The cohomology $H^{even}(X,\mathbf Q)$ is generated by 
$1, J, J^2$ where $J$ is the class dual to a line $C$ in $\mathbf P^2$. 
As shown in the toric diagram in Fig.3, the toric divisor $D$ 
represents the section $\mathbf P^2$. The construction of the secondary 
polytope is easy, and we find $l^{(1)}=(1,1,1,-3)$ for the local 
parameter $x=a^{l^{(1)}}$ of the GKZ hypergeometric system. With these 
data, we may write down the cohomology-valued hypergeometric series, 
\begin{equation}
w(x;\frac{J}{2\pi i})=\sum_{n=0}^{\infty} 
\frac{1}{\Gamma(1+n+\rho)^3\Gamma(1-3(n+\rho))} 
x^{n+\rho} |_{\rho=\frac{J}{2\pi i}}.
\mylabel{eqn:ExZ3I}
\end{equation}
The Taylor series expansion with respect to $\rho$ and the ring structure 
of $H^{even}(X,\mathbf Q)$ define the hypergeometric series of our interest. 
As conjectured in Conjecture \ref{thm:conj3}, the integral and symplectic 
structure of the hypergeometric series comes from $K(X)$ and $K^c(X)$, 
respectively.

\vskip0.5cm
The tautological line bundles form an integral basis of $K(X)$ and, 
following \cite{Re}\cite{Cr}, they are easily determined as  
$$
\mathcal F_0=\mathcal O_X \;,\; 
\mathcal F_1=\mathcal O_X(J) \;,\; 
\mathcal F_2=\mathcal O_X(2J) \;.\; 
$$
We can determine the dual basis $S_k$ following \cite{IN}. Here, instead, we 
fix the D-brane basis $\{\mathcal B_I \}$ to 
$\{ \mathcal O_p, \mathcal O_C(-1), \mathcal O_D(-2) \}$, 
where $\mathcal O_C(-1):=\mathcal O_C \otimes \mathcal O_X(-J)$ 
and $\mathcal O_D(-2):=\mathcal O_D \otimes \mathcal O_X(-2J)$. 
The relations to the bases $S_k$ are easily worked out to be  
\begin{equation}
\mathcal O_p=S_0+S_1+S_2 \;,\;\; 
\mathcal O_C(-1)=S_1+ 2 S_2 \;, \;\; 
\mathcal O_D(-2)=S_2, 
\mylabel{eqn:BbyS}
\end{equation}
and also the symplectic form (\ref{eqn:sympC}) becomes  
\begin{equation}
(\chi(\mathcal B_I,\mathcal B_J))=\left( 
\begin{matrix}
0 & 0 & 0 \\
0 & 0 & -3 \\
0 & 3 & 0 \\
\end{matrix} \right).
\mylabel{eqn:sympZ3}
\end{equation}
From the relations in (\ref{eqn:BbyS}), we find a basis 
$\{ \mathcal F_0, \mathcal F_1-\mathcal F_0, 
\mathcal F_0-2\mathcal F_1+\mathcal F_2 \}$ of $K(X)$, which 
are dual to $\{ \mathcal O_p, \mathcal O_C(-1), \mathcal O_D(-2)\}$. 
Taking $ch: K(X) \rightarrow H^{even}(X,\mathbf Q)$, we then finally 
arrive at a basis $\{ Q_0, Q_2, Q_4 \}$ of $H^{even}(X,\mathbf Q)$, 
$$
Q_0=ch(\mathcal F_0)\;\;,\;\;
Q_2=ch(\mathcal F_1)-ch(\mathcal F_0)\;\;,\;\;
Q_0=ch(\mathcal F_2)-2 ch(\mathcal F_1)+ch(\mathcal F_0)\;\;,\;\;
$$
in which we expand our cohomology-valued hypergeometric series,
$$
w(x;\frac{J}{2\pi i})=
w^{(0)}(x) Q_0 +
w^{(1)}(x) Q_2 +
w^{(2)}(x) Q_4 \;\;.
$$
For convenience, we write explicitly the coefficient hypergeometric series,
\begin{equation}
w^{(0)}(x)=1 \;,\; 
w^{(1)}(x)=\partial_{\tilde\rho} w(x) \;,\; 
w^{(2)}(x)=\frac{1}{2}\partial_{\tilde\rho}^2 w(x)-\frac{1}{2}
\partial_{\tilde\rho}w(x) \;,\; 
\mylabel{eqn:frobSol}
\end{equation}
where $\partial_{\tilde\rho}:=\frac{1}{2\pi i} 
\frac{\partial \;}{\partial\rho}$. 

\vskip0.5cm
Our Conjecture \ref{thm:conj3} predicts that the coefficient hypergeometric 
series $w^{(0)}(x)$, $w^{(1)}(x)$, $w^{(2)}(x)$ thus obtained have 
integral and symplectic monodromy properties with respect to the 
symplectic form (\ref{eqn:sympZ3}). We 
verify these properties in Appendix A. 
Moreover our Conjecture \ref{thm:conj3} predicts that these 
hypergeometric series represents period integrals over the respective 
mirror cycles of 
$\mathcal O_p, \mathcal O_C(-1), \mathcal O_D(-2)$ under 
$mir: K^c(X) \simrightarrow H_3(Y,\mathbf Z)$, 
i.e.,
\begin{equation}
\begin{aligned}
& w^{(0)}(x)=\int_{mir(\mathcal O_p)}\Omega(Y_x) \;\;,\;\; \\
& w^{(1)}(x)=\int_{mir(\mathcal O_C(-1))}\Omega(Y_x) \;\;,\;\;
w^{(2)}(x)=\int_{mir(\mathcal O_D(-2))}\Omega(Y_x) \;\;,\;\;  \\ 
\end{aligned}
\mylabel{eqn:cycleInt}
\end{equation}
where $Y$ is the mirror hypersurface $F(U,V,W_1,W_2)=0$ in 
$\mathbf C^2 \times \mathbf (\mathbf C^*)^2$ defined in (\ref{eqn:mirrorF}). 
In Appendix A, we also construct explicitly the cycles 
which represent $mir(\mathcal O_p), mir(\mathcal O_C(-1))$ 
and $mir(\mathcal O_D(-2))$.  

We remark that, in the case of the two dimensional singularity, 
the hyperk\"ahler rotation helped us identify the mirror cycles 
in $H_2(Y,\mathbf Z)$. In the present case, however, 
it is not so obvious to see that there is a 
canonical (geometric) way to construct the isomorphism 
$mir: K^c(X) \simrightarrow H_3(Y,\mathbf Z)$. Nevertheless, 
our Conjecture \ref{thm:conj3} 
predicts that there is in fact a canonical isomorphism encoded indirectly 
in our cohomology-valued hypergeometric series. This situation is 
parallel to the mirror symmetry of compact Calabi-Yau manifolds, 
where Strominger-Yau-Zaslow construction provides a recipe for the 
canonical mirror correspondence.

\vskip0.5cm
Finally we determine the prepotential. For this, we construct 
the symplectic dual to $\mathcal O_C(-1)$ by 
$-\frac{1}{3} \mathcal O_D(-2)$. Then from (\ref{eqn:cycleInt}) 
the special K\"ahler relation becomes 
$$
(1, t, \frac{\partial F}{\partial t} )=(1, w^{(1)}(x), -\frac{1}{3} w^{(2)}).
$$
Integrating this relation, we obtain the Gromov-Witten invariants of 
the geometry $X=K_{\mathbf P^2}$ (, the first column of Table 1 below). 
It should be noted that the symplectic 
dual $\frac{1}{3} \mathcal O_D(-2)$ is no longer in $K^c(X)$ 
but in $K^c(X) \otimes \mathbf Q$.

\vskip0.5cm

\centerline{\epsfxsize 8truecm\epsfbox{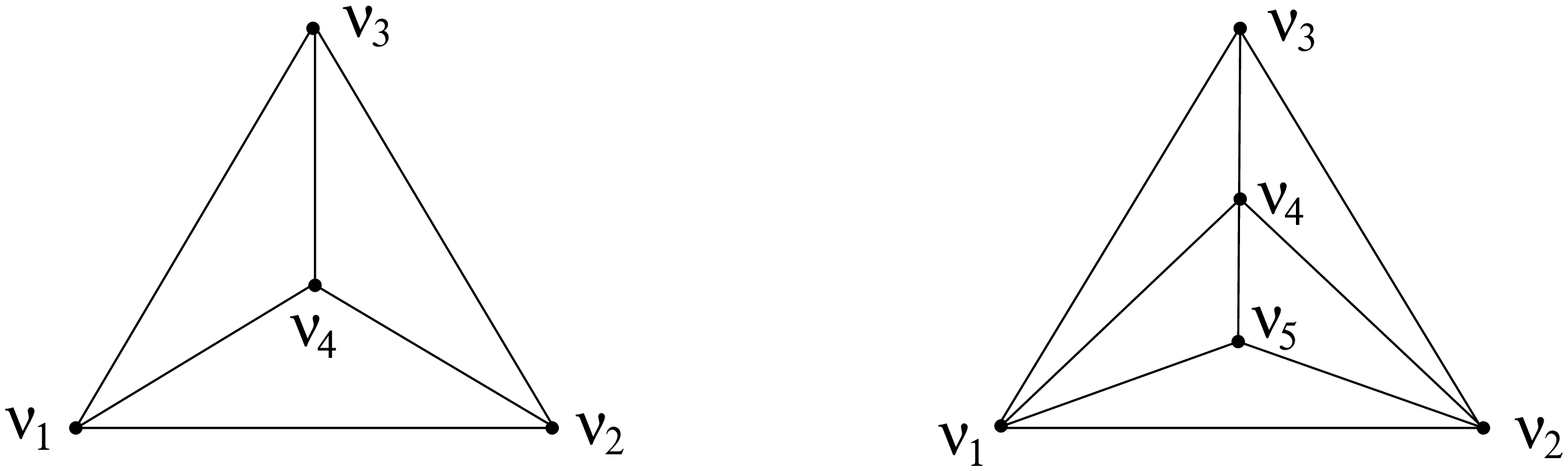}}

{\leftskip1cm \rightskip1cm 
\noindent 
{\sl {\bf Fig.3.} The resolution diagrams of the singularities $\mathbf C^3/G$ 
with $G=\langle \frac{1}{3}(1,1,1) \rangle=\mathbf Z_3$ (left) and 
$G=\langle \frac{1}{5}(1,1,3) \rangle=\mathbf Z_5$ (right).  
\par}}

\vskip0.8cm

\noindent 
{\it Example (2)}: 
As the second example, we briefly sketch the case 
$X=\widehat{\mathbf C^3/\mathbf Z_5}$ with the group action 
specified by the weights $\frac{1}{5}(1,1,3)$. The crepant resolution 
is described in Fig.3. The toric divisors $D_4:=D_{\nu_4}$ and 
$D_5:=D_{\nu_5}$ are identified, respectively, with the Hirzebruch 
surface $\mathbf F_3$ and $\mathbf P^2$. We will denote the rational curves 
that appear as the zero section of $\mathbf F_3$ by $b \equiv C_1$, and also 
the fiber by $f\equiv C_2$. Thus the intersection numbers on $\mathbf F_3$ are 
$b\cdot b=-3$, $b\cdot f=1$, $f\cdot f =0$. $H^2(X,\mathbf Z)$ is generated 
by $J_1$ and $J_2$ which measure the volumes $C_1=b$ and $C_2=f$, 
respectively (, i.e. $\int_{C_i}J_k=\delta_{ik}$). By the standard 
toric method (see, e.g., \cite{Oda2}\cite{Fu}), we  
can determine the ring structure of the cohomology as 
$$
H^{even}(X,\mathbf Q)=\mathbf Q[J_1,J_2]/(J_1^2,J_2^2-3J_1J_2,J_2^3).
$$
The construction of the secondary fan $Sec(\Sigma_G)$ is straightforward 
and we find 
$l^{(1)}=(1,1,0,1,-3)$ and $l^{(2)}=(0,0,1,-2,1)$ for the 
local parameters $x_1=x:=a^{l^{(1)}}$ and $x_2=y:=a^{l^{(2)}}$ of the 
hypergeometric series. With these data we can write 
down the cohomology-valued hypergeometric series (\ref{eqn:covGKZ}) with 
$$
c(n,m)=\frac{1}{\Gamma(1+n)^2\Gamma(1+m)\Gamma(1+n-2m)\Gamma(1-3n+m)} \;\;.
$$

\vskip0.5cm
For the expansion of the cohomology-valued hypergeometric series, we 
first fix a symplectic D-brane basis of $K^c(X)$ by $\{ \mathcal B_J \} 
= \{ \mathcal O_p, \mathcal O_b(-J_1), \mathcal O_f(-J_2), 
\mathcal O_{\mathbf P^2}(-2J_1)$, 
$\mathcal O_{\mathbf F_3}(-2J_2) \}$. 
Then it is straightforward to express these bases in terms of $S_k$ by 
evaluating the pairings of $\mathcal B_J$'s 
with the tautological bundles \cite{Re}\cite{Cr}; 
$$
\mathcal F_0 = \mathcal O_X \;,\;\,
\mathcal F_1 = \mathcal O_X(J_1) \;,\;\,
\mathcal F_2 = \mathcal O_X(2J_1) \;,\;\,
\mathcal F_3 = \mathcal O_X(J_2) \;,\;\,
\mathcal F_4 = \mathcal O_X(J_1+J_2)\;.
$$
Here we summarize the results together with their duals  
under the pairing $K^c(X) \times K(X) \rightarrow \mathbf Z$:
\begin{equation}
\begin{matrix}
& K^c(X) & K(X) \\
\mathcal O_p&=S_0+S_1+S_2 + S_3+S_4 \hfill \quad  
& -\mathcal F_0+2 \mathcal F_1 + 2 \mathcal F_3 -2 \mathcal F_4 \hfill 
& \quad \\ 
\mathcal O_{b}(-J_1)&=S_1+2 S_2+S_4 \hfill  
& -\mathcal F_3+\mathcal F_4 \hfill &\\
\mathcal O_{f}(-J_2)&=S_3+S_4  \hfill  
&\mathcal F_0-2 \mathcal F_1 - \mathcal F_3 +2 \mathcal F_4 \hfill &\\
\mathcal O_{\mathbf P^2}(-2J_1)&=S_2  \hfill  
& \mathcal F_0-2 \mathcal F_1 + \mathcal F_2  \hfill &\\ 
\mathcal O_{\mathbf F_3}(-2J_2)&=2S_0 + S_1  \hfill  
& \mathcal F_0-\mathcal F_1 - \mathcal F_3 + \mathcal F_4 \hfill &
\end{matrix}
\mylabel{eqn:tableSF}
\end{equation}
The symplectic form on $K^c(X)$ may be evaluated using [Corollary 5.3. in 
IN] as 
\begin{equation}
(\chi(\mathcal B_I,\mathcal B_J))=\left( 
\begin{matrix}
0 & 0 & 0 & 0 & 0\\
0 & 0 & 0 & -3 & 1\\
0 & 0 & 0 & 1 & -2\\
0 & 3 & -1 & 0 & 0\\
0 & -1 & 2 & 0 & 0\\
\end{matrix} \right).
\mylabel{eqn:sympZ5}
\end{equation}
Evaluating the Chern characters of the tautological bundles 
in (\ref{eqn:tableSF}), we have for the basis 
$\{ Q_J \}$ of $H^{even}(X,\mathbf Q)$,
$$
Q_0=1-2J_1J_2 \;;\; 
Q_2^1 = J_1+\frac{1}{2}J_1^2+J_1J_2 \;,\;
Q_2^2 = J_2+\frac{7}{2}J_1J_2 \;;\;
Q_4^1 = J_1^2 \;,\; 
Q_4^2 = J_1J_2 \;,
$$
by which we arrange the cohomology-valued hypergeometric series as 
$$
w(x;\frac{J}{2\pi i})=
w^{(0)}(x) Q_0 + 
\sum_{a=1}^2w^{(1)}_a(x) Q_2^a + 
\sum_{b=1}^2w^{(2)}_b(x) Q_4^b \;, 
$$
(where $w^{(0)}(x)\equiv 1$ from the form $c(n,m)$ above).  
Our Conjecture \ref{thm:conj3} predicts that the coefficient 
hypergeometric series $\{ w^{(0)}(x), w^{(1)}_a(x), w^{(2)}_b(x) \}$ 
should be integral and symplectic with respect to the symplectic form 
(\ref{eqn:sympZ5}). 

\vskip0.5cm

As a check of the mirror symmetry, we determine the prepotential. First, we 
write the special K\"ahler relation (\ref{eqn:specialK}), 
$$
(\, 1 \,,\, t_a  \,,\, \frac{\partial F}{\partial t_a}\,) = 
(\, 1 \,,\, w^{(1)}_a(x) \,,\, \sum_b C^{ab} w^{(2)}_b(x) \,)\;,
$$
with the matrix $(C^{ab})=
\left(\begin{smallmatrix} 
-3 & 1 \\ 1 & -2 \\ \end{smallmatrix}\right)^{-1}$.  
From this relation, it is straightforward to obtain 
$$
\begin{aligned}
&\frac{\partial F}{\partial t_1}=
-\frac{1}{5}(\partial_{\tilde\rho_1}^2
+\partial_{\tilde\rho_1}
 \partial_{\tilde\rho_2}
+\frac{3}{2} 
 \partial_{\tilde\rho_2}^2 )w(x)
+\frac{2}{5}t_1+\frac{7}{10}t_2-\frac{2}{5} \;\;, \\
&\frac{\partial F}{\partial t_2}=
-\frac{1}{10}(\partial_{\tilde\rho_1}^2
+6 
 \partial_{\tilde\rho_1}
 \partial_{\tilde\rho_2}
+ 9 
 \partial_{\tilde\rho_2}^2
)w(x)
+\frac{7}{10}t_1+\frac{27}{10}t_2-\frac{6}{5} \;\;, \\
\end{aligned}
$$
where $\partial_{\tilde\rho_k}=\frac{1}{2\pi i} \partial_{\rho_k}$. 
Just as the case for compact Calabi-Yau manifolds, we can integrate 
the special K\"ahler geometry relation with the prepotential
$$
F(t)=\frac{1}{6}(
-\frac{2}{5}t_1^3-\frac{3}{5}t_1^2t_2-\frac{9}{5}t_1t_2^2
-\frac{9}{5}t_2^3)+
\frac{7}{10}t_1t_2+\frac{1}{5}t_1^2+\frac{27}{20}t_2^2 
-\frac{2}{5}t_1-\frac{6}{5}t_2 
+ F_{inst}(t) \;,
$$
where $F_{inst}(t):=\sum_{n,m\geq 0} N(n,m)q_1^nq_2^m$ represents 
the quantum corrections. Here $N(n,m)$ are the genus zero Gromov-Witten 
invariants of the homology class $\beta=n b + m f \in H_2(X,\mathbf Z)$, 
i.e. $N(n,m)=N_0(n b + m f)$ in the standard notation $N_g(\beta)$ 
of Gromov-Witten invariants of genus $g$ in literatures. 
In Table 1, we have listed the integral 
`invariants' called Gopakumar-Vafa invariants $\tilde N(n,m)$ \cite{GV}. 
One can verify that Gromov-Witten invariants coincides with the 
results from topological vertex technique (see \cite{AKMV}). 

\vskip0.5cm
We may observe the similar integrability of the special K\"ahler relation 
for many other examples. 
In compact situations, this integrability (or more specifically the 
existence of the so-called Griffiths-Yukawa coupling) is a consequence 
of the Griffiths transversality (see \cite{St}) for which we use 
an integration over (mirror) manifolds. 
However it should be noted that the existence 
of the prepotential for non-compact cases seems non-trivial, since 
integrating over a non-compact manifolds needs some special care.

$$
\vbox{\offinterlineskip
\hrule
\halign{ \strut 
\vrule#  
& $\,$ \hfil #  \hfil  
&&\vrule#  
& $\;$ \hfil # \hfil 
& \hfil # \hfil  
& \hfil # \hfil  
& \hfil # \hfil 
& \hfil # \hfil  
& \hfil # \hfil  
& \hfil # \hfil  
\cr 
&  \hskip-2pt n \hskip-3pt $\backslash$ \hskip-3pt m \hskip-2pt 
   && 0 & 1 & 2 & 3 & 4 & 5 & 6  &\cr
\noalign{\hrule} 
&0 &&  0 & -2 & 0  &  0 &   0 &   0 &   0 &\cr
&1 &&  3 &  4 & 3 & 5 & 7 & 9 & 11  &\cr
&2 && -6 & -10  & -12 & -12 & -24 & -56 & -140  &\cr
&3 && 27 & 64 & 91 & 108 & 150 & 294 & 675   &\cr 
&4 && -192 & -572 & -980 & -1332 & -1808 & -2982 &  -5992  &\cr
&5 && 1695 &  6076 & 12259 &18912 & 26983 & 42005 &76608 &\cr
&6 && -17064 & -71740 &-166720  &-289440  &-443394  & -689520 &-1192644 &\cr
&7 &&  188454 &909760 &2394779  &4632120  &7665776  & 12254816  & 
       20764870 &\cr
}
\hrule} 
$$
{\leftskip1cm\rightskip1cm\noindent
{\bf Table 1.} Gopakumar-Vafa numbers   
$\tilde N_0(n,m)=\tilde N_0(n b + m f)$ for 
$X=\widehat{\mathbf C^3/\mathbf Z_5}$. $\tilde N_0(n,0)$ coincides with 
the Gopakumar-Vafa numbers for $\widehat{\mathbf C^3/\mathbf Z_3}$. \par} 

\vskip1cm
\section{ {\bf Conclusion and discussions}}

We have studied in detail the cohomology-valued hypergeometric 
series for the case of non-compact Calabi-Yau manifolds. 
Giving an interpretation of $w(x;\frac{J}{2\pi i})$ 
as the central charge formula (Conjecture \ref{thm:conj3}), 
we have provided supporting evidence for the conjecture in the cases of 
$X=\widehat{\mathbf C^2/\mathbf Z_{\mu+1}}$, and also $X=G$-Hilb $\mathbf C^3$ 
with finite abelian subgroups $G=\mathbf Z_3, \mathbf Z_5 
\subset SL(3,\mathbf C)$. We have also 
clarified the structure of the prepotential for the non-compact 
Calabi-Yau geometries $X=G$-Hilb $\mathbf C^3$.

As a byproduct of our study, 
we have found that K.Saito's system of differential 
equations satisfied by the primitive forms may be replaced by a suitable 
(resonant and reducible) GKZ system, whose solutions we may set up 
easily. 

\vskip0.5cm

As addressed after Conjecture \ref{thm:conj}, our cohomology-valued 
hypergeometric series (or the central charge formula) connects two 
different `monodromy' properties, Fourier-Mukai transforms and the 
monodromy transforms of hypergeometric series (Dehn twists in the 
symplectic mapping class group). We see that the latter monodromy property 
arises associated with the discriminant locus in $\mathcal M_{Sec(\Sigma)}$. 
As shown in section 3 for $X=\widehat{\mathbf C^2/\mathbf Z_{\mu+1}}$, 
the discriminant splits into several irreducible components in the 
$q$-coordinate and the monodromy transform around each irreducible 
component may be identified with a suitable twist functor. From these facts, 
it is conceivable that the group of self-equivalences of $D^b(Coh(X))$, 
i.e. Auteq $D^b(X)$, is generated by these `monodromy' transformations up to 
the shift functors, 
$$
\text{Auteq} 
D^b(\widehat{\mathbf C^2/\mathbf Z_{\mu+1}})/\{\; [k]\,|\,k\in \mathbf Z \;\}
= \left< R_1,\cdots,R_\mu, T_{\mathcal O_{C_1}(-1)}, \cdots, T_{\mathcal 
O_{C_\mu}(-1)} \right>\;,
$$
where $R_k$ represents the functor tensoring the tautological 
line bundle $\mathcal F_k$ and $T_{\mathcal O_{C_k}(-1)}$ is the 
Seidel-Thomas twist. (This been proved affirmatively in a recent 
paper by Ishii and Uehara \cite{IU}.)
In Proposition \ref{thm:GKZlogB}, we have seen explicitly 
that these functors are represented by the corresponding actions on 
the roots $\beta_k(a)$'s, i.e., changing the phases of the roots and 
permuting (braiding) of the roots $\beta_k(a)$'s. This simple picture 
should provide us an intuition about the possible forms of the 
self-equivalences in $D Fuk(Y,\beta)$. 

\vskip0.5cm

A similar correspondence between monodromy transforms and Fourier-Mukai 
transforms, and also the canonical mirror isomorphism $mir: K^c(X) 
\simrightarrow H_3(Y,\mathbf Z)$ are contained in our 
Conjecture \ref{thm:conj3} for three dimensional cases 
$X=\widehat{\mathbf C^3/G}=G$-Hilb $\mathbf C^3$ 
\cite{IN}\cite{BKR}\cite{CI}. 
In three dimensions, there can be several Calabi-Yau resolutions 
of the singularity $\mathbf C^3/G$, and correspondingly there 
are many different types of (Fourier-Mukai) transforms connecting 
these different resolutions \cite{BKR}\cite{CI}. It is interesting 
to study the details of the canonical mirror isomorphism $mir$ from 
our cohomology-valued hypergeometric series in such situations.

As we have done for $X=\widehat{\mathbf C^3/\mathbf Z_3}$, 
in some cases with lower dim$H^2(X,\mathbf Q)$, 
we can determine explicitly the integral cycles and their 
monodromy properties to provide a consistency check for our 
Conjecture \ref{thm:conj3}. We can also read off the isomorphism 
$mir: K^c(X) \simrightarrow H_3(Y,\mathbf Z)$ from our 
cohomology-valued hypergeometric series. This isomorphism $mir$ 
should also follow from the geometric framework like the    
Strominger-Yau-Zaslow construction, which provides a canonical 
equivalence $Mir: D^b_c(Coh(X)) \simrightarrow D Fuk(Y,\beta)$.

\vskip0.5cm

Finally, as for the compact Calabi-Yau (hypersurfaces), 
supporting evidence 
for our Conjecture \ref{thm:conj} is reported in \cite{Hos}, especially 
for dimensions one and two. However, for example, the observation 
(\ref{eqn:quintic}) made for quintic hypersurface $X_5$ still 
needs a justification.  Namely we have to work out 
a symplectic basis of the K-group $K(X_5)$ which yields the 
specific forms of the `charges' $1, 
J-\frac{c_1(X_5)J}{12}-\frac{11}{2}\frac{J^2}{5},
\frac{J^2}{5},\frac{J^3}{5}$, and also their mirror cycles in $H_3(X_5^\vee, 
\mathbf Z)$ with their period integrals $w^{(0)}(x), w^{(1)}(x), 
w^{(2)}(x), w^{(3)}(x)$. 
From the SYZ construction \cite{SYZ}\cite{LYZ} we may expect that $K(X_5)$ is 
generated by the structure sheaf $\mathcal O_X$, 
a skyscraper sheaf $\mathcal O_p$, and 
additional sheaves $\mathcal E$ and $\mathcal F$ which, respectively, have  
their support on a divisor and a curve (,i.e. $D4$ and $D2$ branes). 
The cycles which give the period integrals $w^{(0)}(x)$ and $w^{(3)}(x)$ 
are known from the original work \cite{CdOGP}. In fact, claiming these cycles 
to be the mirror images of $\mathcal O_p$ and $\mathcal O_X$, i.e. the 
Lagrangian torus cycle $T^3$ and the section $S^3$ of the fibration, 
respectively, was the starting point of the SYZ construction. 
However we still lack knowledge 
for the remaining cycles and also the corresponding sheaves 
$\mathcal E$ and $\mathcal F$.

\vfill\eject
\begin{appendix}
\section{ \bf Integrating over vanishing cycles }

\noindent
{\bf (A-1) $\mathbf C^2/\mathbf Z_{\mu+1}$. } As a warm up, let us first 
evaluate the period integral (\ref{eqn:Pi2}) of the primitive form.  
We will obtain the central charge $Z_t(S_k)$ in (\ref{eqn:ZSk}) of the 
sheaf $S_k = \mathcal O_{C_i}(-1)$ as the integral over the vanishing 
cycle $L_i=mir(\mathcal O_{C_i}(-1))$. 

\vskip0.5cm
As in Proposition \ref{thm:GKZ2}, we write the 
roots of the polynomial $\psi(W)=a_0+a_1W+\cdots + a_{\mu+1}W^{\mu+1}$ 
by $\beta_0, \cdots, \beta_\mu$. Then for the  
defining equation of the singularity, we have  
$$
f(a,W)+U^2+V^2 = (W-\beta_0)\cdots(W-\beta_{\mu})+U^2+V^2 \;, 
$$
where we set 
$a_{\mu+1}=1$ by scaling the variable $W$. 
Assume that all the roots $\beta_i$ are real and 
$\beta_0<\beta_1<\cdots<\beta_{\mu}$. When  
$f(a,W)\leq 0 $ holds for $\beta_{k-1} \leq W \leq \beta_k$, 
we may construct a vanishing cycle $L_k$ as $L_k=L_k^+ \cup L_k^- \approx 
S^2$ with
$$
L_k^\pm=\{(\pm \sqrt{|f(a,W)|-V^2},V,W)\in \mathbf R^3 \;|\; 
f_\Sigma(a,W)+V^2 \leq 0 , \; 
\beta_{k-1} \leq W \leq \beta_k \;\} \;.
$$            
Similarly, when $f(a,W)\geq 0$ for 
$\beta_{k-1} \leq W \leq \beta_k$, we construct a vanishing cycle 
changing the variables; $U \rightarrow \sqrt{-1}U, V \rightarrow \sqrt{-1} V$. 
For the first case, the evaluation of the period integral over 
$L_k^+$ proceeds as 
\begin{equation}
\begin{aligned}
& \int_{L_k^+} Res_{F_\Sigma=0}  
\left( \frac{dW dU dV}{f(W)+U^2+V^2} \right) 
= \int_{\beta_{k-1}}^{\beta_k} dW 
\int_{-\sqrt{|f(W)|}}^{\sqrt{|f(W)|}} \frac{(\frac{1}{2})\, dV}
{\sqrt{|f(W)|-V^2}}  \\
&
= \frac{1}{2}\int_{\beta_{k-1}}^{\beta_k} dW \;{\rm Sin}^{-1}
\left(\frac{V}{\sqrt{|f(W)|}}\right)
\bigg\vert_{-\sqrt{|f(W)|}}^{\sqrt{|f(W)|}} 
=\frac{\pi}{2} \int_{\beta_{k-1}}^{\beta_k} dW \;\;, \\
\end{aligned}
\mylabel{eqn:A1}
\end{equation}
where $F_\Sigma:=f(W)+U^2+V^2$. 
We have a similar result for $L_k^-$, and in total 
$\pi (\beta_k - \beta_{k-1})$ for the period integral. Since we have 
$\mu$-independent vanishing cycles in this way, we come to 
the well-known results summarized in Proposition \ref{thm:KsaitoSystem}, 
and also Proposition \ref{thm:GKZ2}. In the above evaluation, 
one should note that the period integrals of the primitive form 
measure the volumes of the vanishing cycles.

\vskip0.5cm
The period integral $\Pi_\gamma(a)$ (\ref{eqn:Pi}) 
has a slightly different shape in the integration measure. 
First let us take $\gamma=L_k \times S^1$ 
for the cycle $\gamma \in H_3(\mathbf C^2 \times \mathbf C^*\setminus 
(F_\Sigma=0), \mathbf Z)$ with a loop $S^1$ which encircles the 
hypersurface. Since the integration over $S^1$ produces the holomorphic 
two form $\Omega(Y_x)=\frac{i}{2\pi^2}Res(\frac{1}{f(W)+U^2+V^2}
dUdV\frac{dW}{W})$, we have 
\begin{equation}
\int_{L_k^+} \Omega(Y_x)
= \frac{i}{4\pi^2}\int_{\beta_{k-1}}^{\beta_k} \frac{dW}{W} 
\int_{-\sqrt{|f(W)|}}^{\sqrt{|f(W)|}} \frac{dV}{\sqrt{|f(W)|-V^2}}
=\frac{i}{4\pi} \int_{\beta_{k-1}}^{\beta_k} \frac{dW}{W} \;\;, 
\mylabel{eqn:A2}
\end{equation}
and a similar result for $L_k^-$. Thus we arrive at the central 
charge of the sheaves $S_k=\mathcal O_{C_k}(-1)$ in (\ref{eqn:ZSk}) 
and justifies our claim that $mir(\mathcal O_{C_k}(-1))=L_k$ that 
follows from the cohomology-valued hypergeometric 
series (\ref{eqn:WxJlocal}).  
One should note that the logarithmic 
behaviors of the solutions in Proposition \ref{thm:GKZlogB} come 
from the torus invariant measure $\frac{dW}{W}$ in (\ref{eqn:Pi}).  
The claimed relation $mir(\mathcal O_p)=T_0$ for a cycle 
$T_0\approx S^1\times S^1$ can also be justified by evaluating 
the period integral $\int_{T_0}\Omega(Y_x)=1$ (, see the corresponding 
cycle $T_0$ and the evaluation of the period integral over it in  
three dimensions below).

\vskip0.5cm
\noindent
{\bf (A-2) $\mathbf C^3/\mathbf Z_{3}$.} 
The evaluations of the period integrals in (\ref{eqn:A1}),(\ref{eqn:A2})  
extend to three dimensions. In three dimensions, however, we see that 
some vanishing cycles require careful treatment and in fact become 
non-compact cycles with non-trivial fundamental groups. This non-compactness 
of certain (vanishing) cycles comes from the tori $(\mathbf C^*)^2$ of the 
ambient scape $\mathbf C^2 \times (\mathbf C^*)^2$, and is detected only 
by the invariant measure of the period integral of the local mirror symmetry 
(\ref{eqn:A2}). To illustrate this, we first construct all the cycles which 
reproduce the period integrals (\ref{eqn:cycleInt}). We will then 
contrast the results to the corresponding period integrals of 
the primitive form.

\vskip0.5cm
\noindent
{$\bullet$ \it (Non-compact) Vanishing cycles and period integrals.} 
\vskip0.2cm

Let us start with the toric data for the crepant resolution 
$\widehat{\mathbf C^3/\mathbf Z^3}$, i.e., $\nu_1=(1,0,0), \nu_2=(0,1,0), 
\nu_3=(0,0,1), \nu_4=(1/3,1/3,1/3)$ in $N_G$. We fix an isomorphism 
$\varphi: N_G \cong \mathbf Z^3$ such that $\varphi(\nu_1)=(1,0,0), 
\varphi(\nu_2)=(1,2,1), \varphi(\nu_3)=(1,1,2),$ $\varphi(\nu_4)=(1,1,1)$,  
and consider the following deformation family of hypersurfaces in 
${\mathbf C}^2\times (\mathbf C^*)^2$;
$$
F_\Sigma(a,U,V,W):=a_1+a_2W_1^2W_2+a_3W_1W_2^2+a_4W_1W_2 + U^2 + V^2 =0 \;,
$$
with $(a_1,..,a_4) \in (\mathbf C^*)^4$. The period integral 
of the local mirror symmetry is defined in general in (\ref{eqn:3-dimP}) and 
satisfies the GKZ hypergeometric system (\ref{eqn:3dimGKZ}) in 
Proposition \ref{thm:propGKZ3}. This GKZ hypergeometric system introduces 
the toric compactification ${\mathcal M}_{Sec(\Sigma)}$ in terms of 
the secondary fan $Sec(\Sigma)$ for the deformation parameters $(a_1,..,a_4)$. 
In the present case, we have 
${\mathcal M}_{Sec(\Sigma)} \cong {\mathbf P}^1$ with 
the local parameter $x=\frac{a_1a_2a_3}{a_4^3}$ near the large complex 
structure limit. By scaling the variables, we may set $(a_1,a_2,a_3,a_4)=
(a,1,1,1)$, and hence we have the mirror family $\{Y_x\}$ with 
the period integral parametrized by $x=a$;
\begin{equation}
\int_L \Omega(Y_x) = \frac{1}{4\pi^3} \int_L Res_{F_\Sigma=0} \left( 
\frac{1}{a+f_1(W)+U^2+V^2} dUdV \frac{dW_1}{W_1} \frac{dW_2}{W_2} 
\right) \;,
\mylabel{eqn:ApOmega}
\end{equation}
where $a+f_1(W)=a+W_1W_2(1+W_1+W_2)$ and $L$ is a cycle in 
$H_3(Y_x,\mathbf Z)$. In what follows, we construct integral 
cycles $L$ and evaluate the above period integral.

\vskip0.5cm

First, we construct a torus cycle $T_0\approx S^1 \times S^1 \times S^1$ which 
fixes the normalization of the period integral. To do this, we introduce new 
variables $\tilde U = U+i V, \tilde V=U-i V$ and write the defining equation 
as 
$$
F_\Sigma(a,U,V,W)=a+f_1(W)+U^2+V^2 = a+ f_1(W) + \tilde U \tilde V \;\;. 
$$
We define the cycle to be
$$ 
T_0:=\{ (\tilde U,\tilde V,W_1,W_2) \in \mathbf C^2 \times (\mathbf 
C^*)^2 \;|\; \tilde U=\frac{-a-f_1(W)}{\tilde V}, \; 
|\tilde V|=|W_1|=|W_2|=\varepsilon \} \;.
$$
Then it is straightforward to evaluate $\int_{T_0}\Omega(Y_x)=1$ since  
$$
\frac{1}{4\pi^3} 
Res_{F_\Sigma=0}\left( \frac{(-2 i)^{-1} d\tilde U d \tilde V}{
a+f_1(W)+\tilde U \tilde V}\frac{dW_1}{W_1}\frac{dW_2}{W_2} \right) 
=\frac{1}{(2\pi i)^3} 
\frac{d\tilde V}{\tilde V} 
\frac{dW_1}{W_1}\frac{dW_2}{W_2}  \;. 
$$
This normalizes the period integral and verifies the first relation 
in (\ref{eqn:cycleInt}) with $mir(\mathcal O_p)=T_0$.

\vskip0.5cm
\centerline{\epsfxsize 10.5truecm\epsfbox{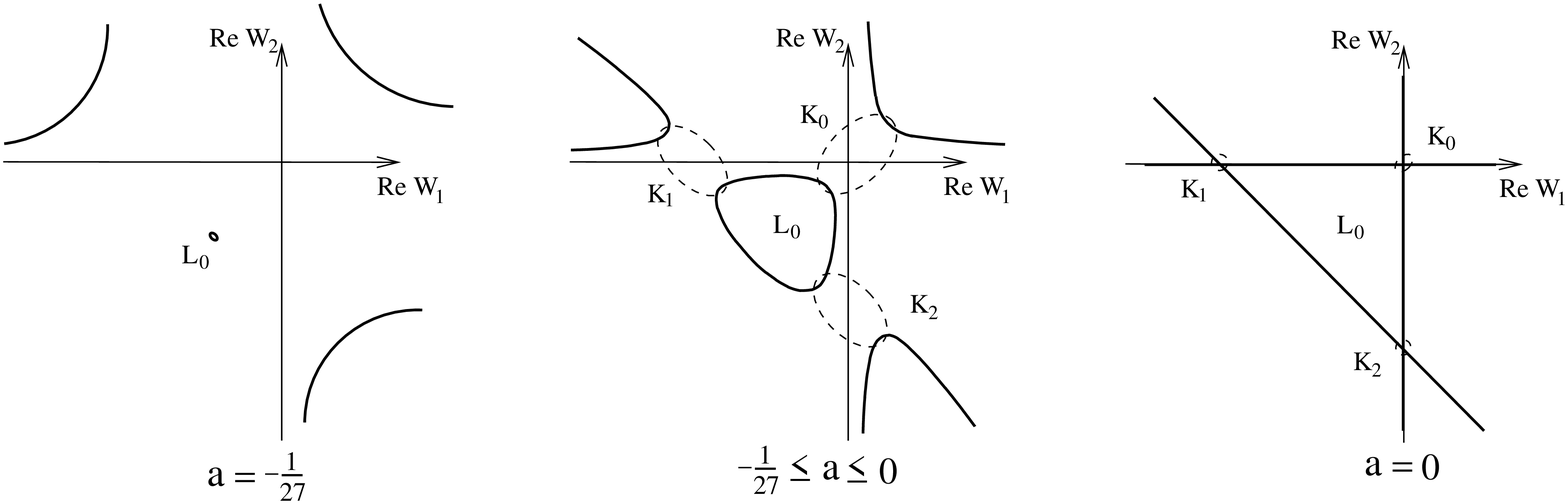}}
{\leftskip1cm \rightskip1cm 
\noindent 
{\sl {\bf Fig.4.} Vanishing cycles near the critical points $(0,0),(-1,0),
(0,-1)$, $(-\frac{1}{3},-\frac{1}{3})$ of the hypersurface 
$a+W_1W_2(1+W_1+W_2)+U^2+V^2=0$ in $\mathbf C^2 \times (\mathbf C^*)^2$ 
$(-\frac{1}{27} \leq a \leq 0)$. 
They are termed $K_0,K_1,K_2$ and $L_0$, respectively, 
and have topologies $K_i \approx S^3\setminus S^1 \, (i=0,1,2)$ and  
$L_0 \approx S^3$.  \par}}

\vskip0.5cm

Secondly, we describe a vanishing cycle which verifies the third 
relation in (\ref{eqn:cycleInt}). Following general arguments on 
singularities, e.g. in \cite{AGV}, we may have vanishing cycles looking 
at the critical points of $f_1(W)$(, more precisely $f_1(W)+U^2+V^2$). 
We see that $f_1(W)=W_1W_2(1+W_1+W_2)$ has four critical points,
$$
(W_1,W_2)_{crit}=(0,0),(-1,0),(0,-1),(-1/3,-1/3), 
$$
with critical values $0, 0, 0, \frac{1}{27}$, respectively. The vanishing 
cycle at $(-1/3,-1/3)$ is easy to describe in a way similar to (A-1). 
The result is $L_0:=L_0^+ \cup L_0^- \approx S^3$ with 
$$
L_0^\pm=\bigg\{ (\sqrt{-1}u,\sqrt{-1}v,W_1,W_2) \,\bigg|\, 
\begin{matrix}
v=\pm \sqrt{a+f_1(W)-u^2}, \;\; a+f_1(W) \geq 0, \\
-\sqrt{a+f_1(W)} \leq u \leq \sqrt{a+f_1(W)} 
\end{matrix} 
\bigg\} \;, 
$$
where $u,v,W_1,W_2$ above are supposed to be real, and also $-\frac{1}{27}
\leq a \leq 0$ (see Fig.4). The evaluation of the 
period integral is quite parallel to (\ref{eqn:A2}), and we have
\begin{equation}
\int_{L_0} \Omega(Y_x)=\frac{1}{4\pi^2} \int_{a+W_1W_2(1+W_1+W_2) \geq 0} 
\frac{dW_1}{W_1}\frac{dW_2}{W_2} \;,
\mylabel{eqn:vsL}
\end{equation}
where $\int_{a+W_1W_2(1+W_1+W_2)\geq 0}$ means the integral over 
the closed domain which appears in the third quadrant of the real $W_1, 
W_2$ plane for $-\frac{1}{27} \leq a \leq 0$ (see Fig.4). 

We may evaluate the integral (\ref{eqn:vsL}) near $a=0$ in a 
straightforward way, by setting $X=W_1, Y=W_2$, as 
$$
\frac{1}{4\pi^2} \int_{Y_{min}(a)}^{Y_{max}(a)} \frac{dY}{Y} 
\int_{X_{min}(a,Y)}^{X_{max}(a,Y)} \frac{dX}{X} 
= \frac{1}{8 \pi^2} (\log(-a))^2 + \cdots ,
$$
where $X_{min}(a,Y) < X_{max}(a,Y)$ are roots of $a+XY(1+X+Y)=0$, and 
$Y_{min}(a),$ $Y_{max}(a)$ are determined from the conditions 
$$
a+XY(1+X+Y)=\frac{\pd\;}{\pd X}(a+XY(1+X+Y))=0
$$ 
and 
$$
Y_{min}(-\frac{1}{27})=Y_{max}(-\frac{1}{27})=-1/3.
$$ 
The behavior 
$\frac{1}{8 \pi^2} (\log(-a))^2 + \cdots$ is easily seen by observing 
$$
\int_{X_{min}(a,Y)}^{X_{max}(a,Y)} \frac{dX}{X} = \log(-a) - \log(-Y(1+Y)^2) 
+ O(a)
$$ 
and 
$$
Y_{min}(a)=-1-2 \sqrt{-a} -2 a + \cdots, \; 
Y_{max}(a)=4a-32 a^2 +\cdots
$$ 
for small $|a|$. Clearly this period integral 
vanishes at $a=-\frac{1}{27}$. This vanishing property and the $(\log(-a))^2$ 
behavior near $a=0$ suffice to identify the period integral with $w^{(2)}(x)
=\frac{1}{2}\pd_{\tilde\rho}^2 w(x) -\frac{1}{2}\pd_{\tilde\rho} w(x)$ 
in (\ref{eqn:frobSol}), which has the same vanishing property at 
$x=-\frac{1}{27}$ and also the expansion near $x=0$, 
$$
w^{(2)}(x)=-\frac{1}{8\pi^2} (\log(-x))^2+\frac{1}{8} 
+ \frac{3}{2 \pi^2} x \log(-x) 
+\frac{9}{4\pi^2}x - \frac{423}{16 \pi^2} x^2 + \cdots \;.
$$
The difference in sign is simply a matter of the orientation of the 
vanishing cycle $L_0$. Thus we verify the third relation in 
(\ref{eqn:cycleInt}) with $mir({\mathcal O}_D(-2))=L_0 \approx S^3$ up to the 
orientation of the cycle.

\vskip0.5cm

Finally, let us look carefully at the vanishing cycles at the critical points 
$(0,0)$, $(-1,0)$, $(0,-1)$, 
which we call $K_0,K_1,K_2$, respectively. Usually 
vanishing cycles  have topology of $S^3$. However this is not the case 
for the cycles $K_i$'s since we consider the mirror geometry in 
${\mathbf C}^2 \times ({\mathbf C}^*)^2$, (see Fig 4). It will turn out 
that they have topology $S^3 \setminus S^1$ and hence non-compact.

Let us first describe the vanishing cycle at the critical point $(0,0)$. 
Similarly to $L_0$, we define $K_0=K_0^+ \cup K_0^-$ using real 
variables $A,B,U,V$ by 
$$
K_0^\pm=\bigg\{ (U,V,A+i B,A-i B) \,\bigg|\, 
\begin{matrix}
U=\pm \sqrt{|a+f_1(A,B)|-V^2}, \;\; a+f_1(A,B) \leq 0, \\
-\sqrt{|a+f_1(A,B)|} \leq V \leq \sqrt{|a+f_1(A,B)|} 
\end{matrix} 
\bigg\} \;, 
$$
where $a+f_1(A,B)=a+(A^2+B^2)(1+2A)$.  
Note that the locus $A=B=0$ is not 
contained in $K_0^\pm$ because $W_1=A+i B, W_2=A-i B$ live 
in $\mathbf C^*$. 
Since the excluded loci $A=B=0$ in $K_0^+$ and $K_0^-$ make up $S^1$, 
we see that $K_0 = K_0^+ \cup K_0^- \approx S^3 \setminus S^1$. 
After some calculations similar to (\ref{eqn:A2}), we obtain
$$
\int_{K_0} \Omega(Y_x) = \frac{1}{4\pi^3} \pi \int_{a+(A^2+B^2)(1+2A) \leq 0} 
\frac{ -2i dA dB}{(A^2+B^2)} \;\;, 
$$
where we mean by $a+(A^2+B^2)(1+2A)\leq 0$ 
the closed domain which appears near the origin in the real $A,B$ plane 
for $-\frac{1}{27} \leq a \leq 0$. 
Note that the integral diverges if the locus $A=B=0$ were 
contained in $K_0$. We may isolate this divergent contribution from 
the origin by introducing the polar coordinate 
$r \cos \theta = \sqrt{1+2A}\, A$, $r\sin \theta = \sqrt{1+2A}\, B$ 
satisfying $r^2 \cos^2 \theta = (1+2 A) A^2$. The calculation proceeds 
as follows;
\begin{equation}
\begin{aligned}
\int_{K_0} \Omega(Y_x)&=
\frac{-i}{2 \pi^2} \int_{\sqrt\epsilon}^{\sqrt{-a}} \frac{dr}{r} 
\int_0^{2\pi} d\theta \frac{1+2 A(r,\theta)}{1+3 A(r,\theta)}  \\
 &= \frac{1}{2\pi i} \frac{1}{3}(-6a+45a-560a^3+\cdots) + \frac{1}{2\pi i} 
(\log(-a) - \log \epsilon) \;,
\end{aligned}
\mylabel{eqn:intK0}
\end{equation}
where $\epsilon >0$ is a small constant to handle the noncompact cycle 
$K_0 \approx S^3\setminus S^1$, 
and the divergent term $\log \epsilon$ 
when $\epsilon \rightarrow 0$ will be ignored in the following. (Here is 
an ambiguity about the cut of the function $\log$. Precisely, we should 
say that we drop the divergent term $\log(-\epsilon)=\log \epsilon - \pi i$.)
This series expansion should be compared with the hypergeometric series
$w^{(1)}(x)=\pd_{\tilde\rho}w(x)$ in (\ref{eqn:frobSol}), which has the 
form near $x=0 \, (x=a)$ 
$$
w^{(1)}(x)=\frac{1}{2\pi i} \log(x) + \frac{1}{2\pi i} ( 
-6 x + 45 x^2 -560 x^3 + \frac{17325}{2} x^4 +\cdots ) \;\;.
$$
The slight difference in the factor $\frac{1}{3}$ can be explained by 
the contribution from the similar non-compact vanishing cycles 
$K_1, K_2$ at the critical points $(-1,0)$ and $(0,-1)$, respectively. 
Let us describe the cycle $K_1=K_1^+ \cup K_1^-$ by 
$$
K_1^\pm=\bigg\{ (U,V,-1-2 A, A+i B) \,\bigg|\, 
\begin{matrix}
U=\pm \sqrt{|a+f_1(A,B)-V^2|}, \;\; a+f_1(A,B) \leq 0, \\
-\sqrt{|a+f_1(A,B)|} \leq V \leq \sqrt{|a+f_1(A,B)|} 
\end{matrix} 
\bigg\} \;. 
$$
Here similarly to $K_0$, $a+f_1(A,B) = a+(A^2+B^2)(1+2A)$ and we mean 
by $a+f_1(A,B) \leq 0$ to represent the closed domain near the origin  
in the real $A,B$ plane which appears for $-\frac{1}{27} \leq a \leq 0$. 
As is the case for $K_0$, the locus 
$A=B=0$ is excluded and hence we have $K_1 \approx S^3 \setminus S^1$. 
The integration over this non-compact cycle $K_1$ may be done by 
changing variables to the polar coordinates. After some manipulations, we 
get a similar result to (\ref{eqn:intK0});
\begin{equation}
\begin{aligned}
\int_{K_0} \Omega(Y_x) 
&=\frac{-i}{2 \pi^2} \int_{\sqrt\epsilon}^{\sqrt{-a}} \frac{dr}{r} 
\int_0^{2\pi} d\theta \frac{-A(r,\theta)}{1+3 A(r,\theta)}  \\
 &= \frac{1}{2\pi i} \frac{1}{3}(-6a+45a-560a^3+\cdots) \;, 
\end{aligned}
\mylabel{eqn:intK1}
\end{equation}
which does not contain the divergent term $\log(-a)-\log \epsilon$. 
We should note 
in the difference between (\ref{eqn:intK0}) and (\ref{eqn:intK1}) 
that $K_0$ is {\it not} homologous to $K_1$ in $H_3(Y_x,\mathbf Z)$ although 
they have the same topology $S^3 \setminus S^1$.  

By symmetry under $W_1 \leftrightarrow W_2$, we have a similar construction  
of the cycle $K_2=K_2^+ \cup K_2^- \approx S^3 \setminus S^1$ and obtain 
the same result as above for $\int_{K_2} \Omega(Y_x)$. 
Putting all the above results together, we obtain
$\int_{K_0+K_1+K_2} \Omega(Y_x)  = w^{(1)}(x)$,
which entails $mir(\mathcal O_C(-1))=K_0+K_1+K_2$ from (\ref{eqn:cycleInt}).

\vskip0.3cm

To summarize, we read the mirror isomorphism 
$mir: K^c(X) \simrightarrow H_3(Y,\mathbf Z)$ from the first part of our 
Conjecture \ref{thm:conj3} (and the equation (\ref{eqn:cycleInt})), 
up to orientations of cycles, as
\begin{equation}
mir(\mathcal O_p)=T_0  \;\;,\;\;
mir(\mathcal O_C(-1))=K_0+K_1+K_2 \;\;,\;\;
mir(\mathcal O_D(-2)) = L_0 \;,
\mylabel{eqn:summaryMir}
\end{equation}
where the topology of the cycles are $T_0 \approx S^1 \times S^1 
\times S^1$, $K_i \approx S^3\setminus S^1 (i=0,1,2)$ and 
$L_0 \approx S^3$.

\vskip0.5cm
\noindent
{$\bullet$ \it Monodromy and the symplectic form.} 
\vskip0.2cm

We have identified the period integrals over the cycles with the 
hypergeometric series by   
$\int_{T_0} \Omega(Y_x)=w^{(0)}(x)$, 
$\int_{K_0+K_1+K_2} \Omega(Y_x)=w^{(1)}(x)$, 
$\int_{L_0} \Omega(Y_x)=w^{(2)}(x)$, where 
$(w^{(0)}(x),w^{(1)}(x),w^{(2)}(x))=
(1, \pd_{\tilde\rho}w(x), \frac{1}{2}\pd_{\tilde\rho}^2 w(x) - 
\frac{1}{2} \pd_{\tilde\rho}w(x) )$. 
The calculation of the monodromy of the hypergeometric series 
$\,^t (w^{(0)}(x),w^{(1)}(x),w^{(2)}(x))$ 
is straightforward (see e.g. \cite{Hos}), and we obtain the monodromy 
matrices about $x=0, x=-\frac{1}{27}, x=\infty$, respectively, as 
$$
M_0=
\left( \begin{matrix}
1 & 0 & 0\cr
1 & 1 & 0\cr 
0 & 1 & 1 \cr \end{matrix} \right)\;,\;\;
M_{1}=
\left( \begin{matrix}
1 & 0 & 0  \\
0 & 1 & -3 \cr
0 & 0 & 1 \cr \end{matrix} \right)\;,\;\;
M_{\infty}=
\left( \begin{matrix}
1  & 0 & 0 \\
-1 & 1 &   3 \cr
 1  & -1 & -2 \cr \end{matrix} \right)\;, \;\;
$$
where $M_\infty:=(M_1M_0)^{-1}$. Now we may verify the 
second part of Conjecture 
\ref{thm:conj3} by observing the equality of the two symplectic forms,
$$
(\chi(\mathcal B_I,\mathcal B_J))=(\# L_I \cap L_J ) = 
\left( \begin{matrix} 0 & 0 & 0 \\ 0 & 0 & -3 \\ 0 & 3 & 0 \\ 
\end{matrix}\right) \;,
$$
for the basis 
$\{\mathcal B_I\}=\{\mathcal O_p, \mathcal O_C(-1), \mathcal O_D(-2)\}$ 
of $K^c(X)$ and its mirror image $\{L_I\} =\{T_0, K_0+K_1+K_2,L_0 \}$ 
in $H_3(Y,\mathbf Z)$. 
The intersection numbers among the cycles $L_I$'s are easily determined 
from their definitions(, see also Fig.4,) under suitable orientations. 
Now the second part of Conjecture \ref{thm:conj3} follows from 
the Picard-Lefschetz theory of the cycles $L_I$. 
By the mirror isomorphism $mir:K^c(X) \simrightarrow H_3(Y,\mathbf Z)$, 
it is conjectured that the cycles $T_0, K_0+K_1+K_2,L_0$ 
constitute an integral basis of $H_3(Y,\mathbf Z)$.

\vskip0.5cm
\noindent
{$\bullet$ \it Period integral of the primitive form.} 
\vskip0.2cm

Here to have things to be contrasted, let us evaluate the period 
integrals of the primitive form,
\begin{equation}
\int_L \mathcal U(a) = \int_L Res_{F_\Sigma=0} \left( 
\frac{dU dV dW_1 dW_2}{a+f_1(W)+U^2+V^2} \right)
\mylabel{eqn:primFm}
\end{equation}
for the cycles $L=T_0, K_0,K_1,K_2,L_0$ which are considered in 
the hypersurface $a+f_1(W)+U^2+V^2=0$ in $\mathbf C^4$. Firstly, 
it is rather clear that the homology class $T_0$ is trivial 
and to have $\int_{T_0} \mathcal U(a)=0$. Also all the cycles 
$K_0,K_1,K_2$ and $L_0$ are simply vanishing cycles having topology of 
$S^3$. The evaluations of the period integral are similar to the 
previous case of the local mirror symmetry, and it is straightforward 
to obtain 
$$
\int_{L_0} \mathcal U(a) = \pi \int_{a+f_1(W)\geq 0} dW_1 dW_2 
=\frac{2 \pi^2}{9 \sqrt{3}}(1+27 a) \,_2F_1(\frac{1}{3},\frac{2}{3},2,1+27a),
$$
where $f_1(W)=W_1W_2(1+W_1+W_2)$ and, as before, $a+f_1(W)\geq 0$ 
represents the closed domain in the third quadrant in the real 
$W_1,W_2$ plane for $-\frac{1}{27} \leq a \leq 0$. For the cycles 
$K_j \,(i=0,1,2)$, we have the same results,
$$
\int_{K_j}\mathcal U(a) = (-2i)\pi \int_{a+f_1(A,B)\leq 0} dA dB 
= 2 \pi^2 i \; a \,_2 F_1(\frac{1}{3},\frac{2}{3},2,-27 a) \;\;,
$$
where $f_1(A,B)=(A^2+B^2)(1+2A)$ and $a+f_1(A,B) \leq 0$ represents 
a closed domain near the origin of the real $A,B$ plane which exists 
for $-\frac{1}{27} \leq a \leq 0$. The result above indicates that 
the cycles $K_0,K_1,K_2$ are homologically equivalent. In fact, we may 
verify the following Picard-Fuchs equation of second order 
satisfied by the period integral (\ref{eqn:primFm}); 
\begin{equation}
\{ \theta_a(\theta_a-1) +3a(3\theta_a-2)(3\theta_a-1) \} 
\int_L \mathcal U(a) = 0 \;\; ,
\mylabel{eqn:2ndPF}
\end{equation}
where $\theta_a=a \frac{\pd \; }{\pd a}$. This equation indicates that 
there exit only two homologically independent cycles.  
In appendix B, we will derive this Picard-Fuchs equation starting 
from the GKZ system appearing in Remark after Proposition \ref{thm:propGKZ3}. 
This equation should be compared with the differential equation 
\begin{equation}
\{\theta_x^2+3x(3\theta_x+2)(3\theta_x+1)\}\theta_x \int_L \Omega(Y_x) =0 \;\;,
\mylabel{eqn:3rdPF}
\end{equation} 
where $\theta_x=x \frac{\pd \; }{\pd x}$. 
The latter third order differential equation, which follows also 
from the GKZ system in Proposition \ref{thm:propGKZ3},  
characterizes the period 
integrals over the cycles (\ref{eqn:summaryMir}).

\vfill\eject
\section{\bf Differential equation (\ref{eqn:2ndPF})}

Here we derive the second order differential equation (\ref{eqn:2ndPF}) 
form the GKZ system for the primitive form (cf.(\ref{eqn:3dimGKZ})) 
observing a factorization of a first order differential operator. 
The same second order differential equation 
can be derived from the K.Saito's system.

As in (A-2), we fix an isomorphism $\varphi: N_G \simrightarrow 
\mathbf Z^3$  so that 
$$
(\varphi(\nu_1) \; \varphi(\nu_2) \; \varphi(\nu_3) \; \varphi(\nu_4)) 
= \left( 
\begin{matrix}
1 & 1 & 1 & 1 \\
0 & 2 & 1 & 1 \\
0 & 1 & 2 & 1 \\
\end{matrix} \right) \;.
$$
From this, we may write the linear operators $\mathcal Z_i'$;
$$
\mathcal Z_1' = \theta_{1}+\theta_{2}+\theta_{3}+\theta_{4} \;,\;
\mathcal Z_2' = 2 \theta_{2}+ \theta_{3}+\theta_{4} +1 \;,\;
\mathcal Z_3' = 3 \theta_{3}+\theta_{4} +1 .
$$
Since the period integral $\int_L \mathcal U(a)$ is annihilated by 
these operators, we have the following relations 
\begin{equation}
\theta_{2}=\theta_{1}-1 \;,\; 
\theta_{3}=\theta_{1}-1 \;,\; 
\theta_{4}=-3\theta_{1}+2 \;,
\mylabel{eqn:ak}
\end{equation}
when acting on $\int_L \mathcal U(a)$. Now consider the operator 
$\Box_l$ for $l=(1,1,1,-3)$. We have 
$$
a_1a_2a_3 \Box_l \int_L \mathcal U(a) = 
\left( \theta_{1}\theta_{2}\theta_{3} - 
\frac{a_1a_2a_3}{a_4^3} (\theta_{4}-2)(\theta_{4}-1)\theta_{4} \right)
\int_L\mathcal U(a) =0 \;. 
$$
Then it is easy to observe the following factorization of the operator 
when we eliminate $\theta_{a_2},\theta_{a_3},\theta_{a_4}$ by (\ref{eqn:ak});
$$
(\theta_{1}-1)\left\{ 
\theta_{1}(\theta_{1}-1)+3 \frac{a_1a_2a_3}{a_4^3} 
(3\theta_{1}-2)(3\theta_{1}-1) \right\} 
\int_C \mathcal U(a) =0 \;.
$$
The inhomogeneous terms (i.e. constants) in (\ref{eqn:ak}) may be 
removed if we define $\Pi_L(a):=\frac{a_2a_3}{a_4^2} 
\int_L \mathcal U(a)$. Even in that case, the shape of the 
above factorized differential operator does not change. 
The resulting homogeneous linear operators imply that  
$\Pi_L(a)=\Pi_L(x)$ with $x=\frac{a_1a_2a_3}{a_4^3}$, and 
enable  us to set $a_1=a,a_2=a_3=a_4=1$ in which we have 
$\Pi_L(x)=\int_L \mathcal U(a)$ with $x=a$. 
Now we arrive at the second order 
differential equation, 
$$
\left\{\theta_a(\theta_{a}-1)+3 a (3\theta_{a}-2)(3\theta_{a}-1) \right\}
\Pi_L(a)=0 \;\;,
$$
by taking the irreducible part. This completes our derivation of 
(\ref{eqn:2ndPF}). 

Similar factorization property of the GKZ systems for 
primitive forms may be observed for $G$ such that $\mathbf C^3/G$ 
has an isolated singularity. This factorization property is 
reminiscent the factorizations observed in \cite{HKTY1} for 
the (extended) GKZ systems which appear in applications of 
mirror symmetry, although the irreducible part shows slightly 
different degenerations as we see above (cf. (\ref{eqn:3rdPF})).

\end{appendix}

\vfill\eject
\vskip1cm

\end{document}